\documentclass[fleqn,usenatbib]{mnras}
\usepackage{amsmath}	% Advanced maths commands
\usepackage{amssymb}	% Extra maths symbols

\usepackage{mathptmx}
\usepackage{txfonts}

\usepackage[T1]{fontenc}
\usepackage{graphicx}	% Including figure files

\newcommand \msun {M$_\odot$}
\newcommand \rsun {R$_\odot$}
\newcommand \drvm  {$\Delta{\rm RV}_{\rm max}$}
\newcommand \fb {$f_{\rm bin}$}
\newcommand \kms {\,km\,s$^{-1}$}

\defcitealias{Maoz_2012}{M12}
\defcitealias{Badenes_2012}{BM12}

\title[Double-WD fraction, separation, and mergers from SPY]{The binary fraction, separation distribution, and merger rate of white dwarfs from SPY}

\author[D. Maoz and N. Hallakoun]{
Dan Maoz$^{1}$\thanks{E-mail: maoz@astro.tau.ac.il (DM)}
and Na'ama Hallakoun$^{1,2}$
\\
$^{1}$School of Physics and Astronomy, Tel-Aviv University, Tel-Aviv 6997801, Israel\\
$^{2}$European Southern Observatory, Karl-Schwarzschild-Stra{\ss}e 2, D-85748 Garching, Germany
}

\date{Accepted XXX. Received YYY; in original form ZZZ}

\pubyear{2016}

\begin{document}
\label{firstpage}
\pagerange{\pageref{firstpage}--\pageref{lastpage}}\maketitle
\begin{abstract}
From a sample of spectra of 439 white dwarfs (WDs) from the ESO-VLT Supernova-Ia Progenitor surveY (SPY), we measure the maximal changes in radial-velocity (\drvm) between epochs (generally two epochs, separated by up to 470\,d), and model the observed \drvm\ statistics via Monte-Carlo simulations, to constrain the population characteristics of double WDs (DWDs). The DWD fraction among WDs is \fb$=0.100 \pm 0.020$ (1$\sigma$, random) $+ 0.02$~(systematic), in the separation range $\lesssim 4$\,AU within which the data are sensitive to binarity. Assuming the distribution of binary separation, $a$, is a power-law, $dN/da\propto a^\alpha$, at the end of the last common-envelope phase and the start of solely gravitational-wave-driven binary evolution, the constraint by the data is $\alpha=-1.3 \pm 0.2$ ($1\sigma$) $\pm 0.2$~(systematic). If these parameters extend to small separations, the implied Galactic WD merger rate per unit stellar mass is $R_{\rm merge}=\left(1-80\right)\times 10^{-13}$\,yr$^{-1}\,M_\odot^{-1}$ ($2\sigma$), with a likelihood-weighted mean of $R_{\rm merge}=(7 \pm 2) \times 10^{-13}$\,yr$^{-1}\,M_\odot^{-1}$ ($1\sigma$). The Milky Way's specific Type-Ia supernova (SN Ia) rate is likely $R_{\rm Ia}\approx 1.1\times 10^{-13}$\,yr$^{-1}\,M_\odot^{-1}$ and therefore, in terms of rates, a possibly small fraction of all merging DWDs  (e.g. those with massive-enough primary WDs) could suffice to produce most or all SNe Ia.
\end{abstract}


%%%%%%%%%%%%%%%%%%%%%%%%%%%%%%%%%%%%%%%%%%%%%%%%%%

%%%%%%%%%%%%%%%%% BODY OF PAPER %%%%%%%%%%%%%%%%%%
\section{Introduction}
\label{sec:Intro}

A large fraction of all stars, and a majority of intermediate-mass and massive stars, are in multiple systems. Multiplicity is an outcome of star formation and early stellar evolution, and thus serves as a probe of those poorly understood processes. Furthermore, binary, triple, and higher-order systems provide the settings for a rich variety of astrophysical phenomena, including interacting, accreting, and merging binaries, various types of supernovae, and gravitational wave sources. However, the demographics of stellar multiplicity are still poorly known, i.e. the distribution of multiplicity index (single, binary, triple...), separation, component mass ratio, and eccentricity, all as a function of stellar mass, age, metallicity, and Galactic environment \citep[e.g.,][]{Duquennoy_1991,Raghavan_2010}. These demographics must be physically linked, at some level, to those of sub-stellar companions -- brown dwarfs (BDs) and planets -- for which our knowledge is even sketchier, and also to those of stellar remnants -- white dwarfs (WDs), neutron stars, and black holes. 

Binarity in WDs is particularly interesting. WDs are the end state of 95\% of all stars, and they are the current state of the majority of all stars ever formed with mass above $1.2$\,\msun. As such, binary WDs provide a fossil probe of the initial binary populations and of their subsequent binary evolution. Systems consisting of close double WDs (DWDs) are potential progenitors of Type-Ia supernovae \citep[SNe Ia; e.g.][]{Maoz_2014}, AM Canum Venaticorum systems (a WD accreting from another degenerate or semi-degenerate companion star), and R Corona Borealis stars \citep[highly magnetic WDs postulated to result from WD mergers; e.g.][]{Longland_2011}. DWDs will be the main foreground of space-based gravitational-wave detectors such as {\it LISA}, both as resolved sources at higher gravitational-wave frequencies, and as an unresolved continuum at lower frequencies. Identifying the individual nearby DWD systems and measuring the binary parameter distribution as a whole for DWDs is therefore important for the budding field of gravitational-wave astronomy. Like other double-compact-remnant binaries (including neutron stars and black holes), DWDs are physically simple and ``clean'' systems in which the evolution of each WD is decoupled from the other WD and driven mainly by cooling via thermal emission from the surface, while the binary evolution is dictated solely by gravitational wave emission (except in the very final merger phases).

Systematic searches for DWDs began in the 1980s \citep[see][and references therein]{Napiwotzki_2004}. There are now over 90 individual close DWD systems for which orbital parameters have been derived \citep[see, e.g.,][and references therein]{Nelemans_2005, Marsh_2011, Debes_2015, Hallakoun_2016, Brown_2016}. Excluding systems with extremely low-mass (ELM) WDs of $\sim 0.2$\,\msun, which are found to be always in binaries, there are about 30 DWD systems with orbital parameters. The statistics of the short-orbit DWD population as a whole were examined by \citet{Maxted_1999}, who studied a sample of 46 WDs and estimated a binary fraction between 1.7\% and 19\%. More recently, \citet[][hereafter M12]{Maoz_2012} developed a statistical method to characterize the DWD population in a sample of WDs, by measuring the distribution of \drvm, the maximum radial velocity (RV) shift between several epochs (two or more) of the same WD. \citetalias{Maoz_2012} showed that even with just two epochs per WD and with noisy RV measurements, the \drvm\ distribution can set meaningful constraints on the binary fraction of the population, \fb, and on the distribution of binary separations, $dN/da$. The \drvm\ method applied to such data permits also an estimate of the merger rate of the DWD population. Statistical inference  about the DWD population is thus possible without follow-up observations and full binary parameter solutions for candidates.

\citet[][hereafter BM12]{Badenes_2012} measured few-epoch RVs in the spectra of $\sim 4000$ WDs from the Sloan Digital Sky Survey (SDSS) and applied the method to the observed \drvm\ distribution. The data constrained \fb\ at separations $a<0.05$\,AU to a $1\sigma$ range of $3-20\%$. Assuming a power-law separation distribution, $dN/da\propto a^\alpha$, at the time of DWD formation (from hence the binary separation evolves solely via gravitational-wave emission), \citetalias{Maoz_2012} and \citetalias{Badenes_2012} showed that $\alpha$ is constrained to the range $-2$ to $+1$, with strong covariance between \fb\ and $\alpha$ (low \fb\ together with a steep negative power-law slope $\alpha$, or a higher \fb\ together with a shallower $\alpha$, can both populate the small separation range with DWD systems and produce the high-\drvm\ tail in the observed distribution). As every combination of \fb\ and $\alpha$ translates to a WD merger rate, \citetalias{Badenes_2012} showed that the WD merger rate per unit stellar mass in the Milky Way is constrained to $R_{\rm merge}=1.4^{+3.4}_{-1}\times 10^{-13}{\rm yr}^{-1}\,M_\odot^{-1}$, a range that straddles the Galactic SN Ia rate per unit stellar mass, $R_{\rm Ia}\approx 1.1\times 10^{-13}{\rm yr}^{-1}\,M_\odot^{-1}$. (The Milky Way's specific SN Ia rate can be reliably estimated from its approximate mass and from the fact that it is an Sbc galaxy; see \citetalias{Badenes_2012}).

The SDSS sample of WDs analysed by \citetalias{Maoz_2012} and \citetalias{Badenes_2012}, while large, suffered from a low RV precision of $\sim 80$\kms, a result mainly of the low resolution of the SDSS spectra and of the fact that most WDs have only a few, highly Stark-broadened hydrogen Balmer absorption lines in their spectra. As a consequence, only the systems in the sample with observed \drvm$\gtrsim 250$\kms (some 15 in number) drove the statistical conclusions. This lower limit in the significantly detectable \drvm\ translated, for typical WD masses, to an upper limit in the binary separation that is probed by the study, of $a\sim 0.05$\,AU. Furthermore, the small number of systems driving the signal results in large statistical errors and in the strong degeneracy between the model parameters.

In the present paper, we apply the method of \citetalias{Maoz_2012} and \citetalias{Badenes_2012} to another sample of multi-epoch WD spectra, from the European Southern Observatory (ESO), 8\,m Very Large Telescope (VLT), Supernova-Ia Progenitor surveY \citep[SPY;][]{Napiwotzki_2001}. SPY was a few-epoch spectroscopic survey of $\sim$800 bright ($V\sim 16$\,mag) WDs, with the objective of using RV differences between epochs to identify close DWD systems that will merge within a Hubble time, thus being potential SN Ia progenitors. Published results from SPY relevant to DWDs include \citet{Napiwotzki_2002} (discovery of a DWD with a mass close to the Chandrasekhar limit); \citet{Karl_2003a,Karl_2003b} (follow-up analysis of several DWDs); \citet{Nelemans_2005} (follow-up analysis of five DWDs from SPY); and \citet{Koester_2009} (catalogue and spectroscopic analysis of hydrogen-dominated WDs, including a list of DWDs). A statistical analysis of the SPY dataset as a whole, and its implications for the binary WD population, has not been published to date.

Out of the full SPY dataset, we select about 500 WDs suitable for our present analysis. Although the sample size is an order of magnitude smaller than the SDSS WD sample of \citetalias{Badenes_2012}, the high spectral resolution and signal-to-noise ratio (S/N) possible with the VLT permit resolving the narrow non-local-thermodynamic equilibrium (NLTE) core of the H$\alpha$ line (and sometimes H$\beta$) that exists in the spectra of DA-type WDs (i.e. WDs with only hydrogen lines in their optical spectra, which constitute the majority of WDs). This provides a typical RV resolution of $1-2$\kms\ per epoch, a factor $\sim 50$ times better than for the SDSS sample. This RV resolution, combined with the distribution of time separations between epochs in SPY, means that the SPY sample is sensitive to DWDs out to separations $a\sim 4$\,AU (see Section~\ref{sec:monte-carlo-simul}, below). While, in principle, the lowest-separation/highest-RV systems are also detectable in SPY, the small sample size of SPY makes it unlikely to ``catch'' those systems, and in fact the largest \drvm\ that we measure is $240$\kms. The SPY sample thus nicely complements the SDSS sample, in so much as it probes the WD population's binarity in the $a=0.05-4$\,AU interval range, compared to the $a=0.001-0.05$\,AU range probed by SDSS (the lower limit in SDSS arising from the exposure length of $\sim 15$\,min, which prevents the detection of RV variations in systems with orbital periods comparable to this time.) The logarithmic interval in separation probed, in principle, by SPY, $a=0.001-4$\,AU, is 2.1 times larger than the $a=0.001-0.05$\,AU logarithmic interval of SDSS. Therefore, for example, for a separation distribution that has equal numbers of binaries per logarithmic interval, one would expect to find a binarity fraction about twice as high in SPY as in SDSS.

From analysis of the SPY sample, below, we find values of \fb, $\alpha$, and the merger rate of the WD population, that are consistent with the findings of \citetalias{Badenes_2012} for the SDSS sample, but are now more tightly constrained. The allowed values of $\alpha$ and $R_{\rm merge}$ strengthen the case for the ``double-degenerate'' progenitor scenario of SNe Ia.

\section{WD sample and RV measurement}
\label{sec:Sample}

\begin{figure*}
\centering
\includegraphics[width=\textwidth]{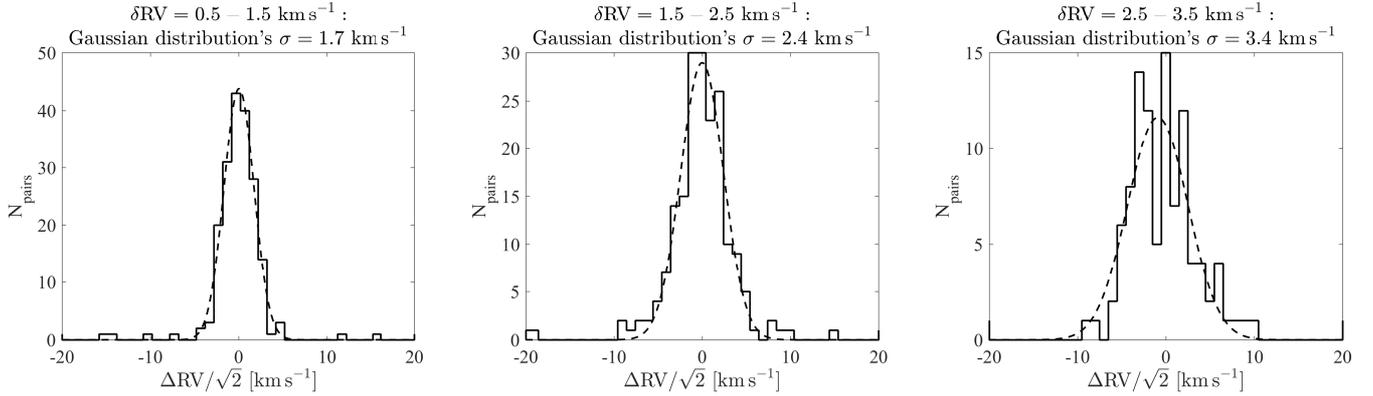}
\caption{Distributions of observed $\Delta$RV differences between two epochs of observations of the same WDs, scaled down by $\sqrt2$, and Gaussian fits to the distributions (dashed), for pairs of measurements having pair-averaged formal RV fitting errors of 0.5-1.5\kms\ (left), 1.5-2.5\kms\ (center), and 2.5-3.5\kms\ (right). Except for some outlier points, resulting from real DWD systems, the distributions appear Gaussian with a $\sigma$ close to the expected value, indicating that the formal RV errors from the Balmer line-profile fitting are reliable.}
\label{fig:RVpairs}
\end{figure*}

\begin{figure*}
\centering
\includegraphics[width=\textwidth]{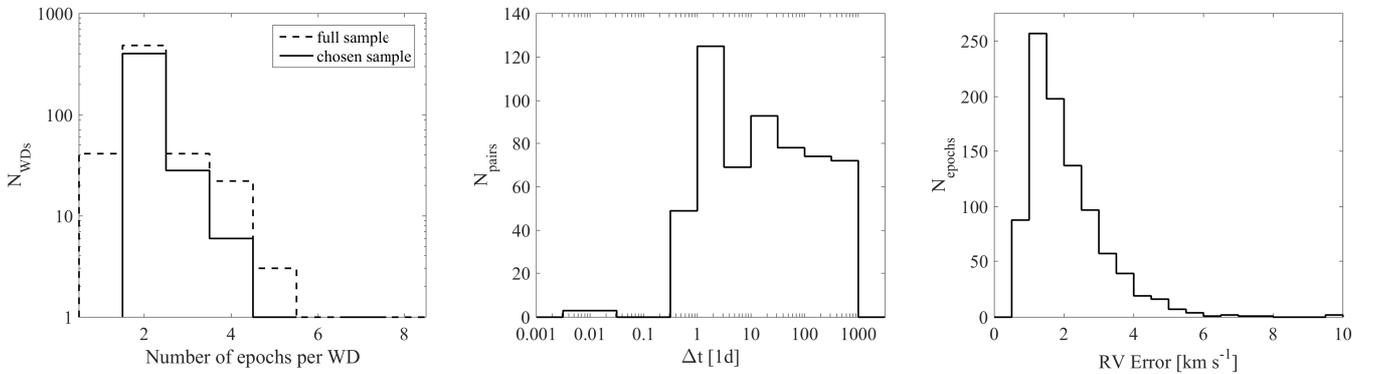}
\caption{Left: the distribution of the number of epochs per WD, for all SPY DA WDs (dashed line) and for our final (solid line) sample. Centre: the distribution of time differences between epochs, for all epochs in the final sample. Right: the distribution of RV errors in the sample.}
\label{fig:EpochDist}
\end{figure*}

The full SPY sample includes some 2200 spectra of about 800 WDs that were observed between the years 2001 and 2003 with the UV-Visual Echelle Spectrograph (UVES) of the ESO Very Large Telescope (VLT). The SPY program setup used UVES in a dichroic mode, covering most of the range between $3200\,\AA$ and $6650\,\AA$, with two $\sim 80\,\AA$ gaps around $4580\,\AA$ and $5640\,\AA$ \citep{Napiwotzki_2003}. The spectral resolution is at least $R=18500$ ($0.36\,\AA$ at H$\alpha$). A S/N per binned pixel ($0.03\,\AA$) of $\gtrsim 15$ was achieved using $5-10$\,min exposures. The SPY WDs were selected from a number of WD compilations and catalogues: the \citet{McCook_1999} catalogue of spectroscopically confirmed WDs; the Hamburg-ESO survey \citep[HES;][]{Wisotzki_1996, Christlieb_2001}; the Hamburg-Quasar survey \citep[HQS;][]{Hagen_1995}; the Montreal-Cambridge-Tololo survey \citep[MCT;][]{Demers_1990, Lamontagne_2000}; and the Edingurgh-Cape survey \citep[EC;][]{Kilkenny_1991}. The SPY targets were selected to have $B \leq 16.5$\,mag and declination $\delta \leq +25^\circ$. Each WD was observed on several epochs (typically two, although in many cases one of the two epochs has noisy data, leaving effectively just one epoch, from which it is impossible to find RV changes, see below). The $\sim 800$ SPY WDs include 615 WDs of type DA \citep[with spectra dominated by hydrogen Balmer absorption lines, see][]{Koester_2009}, 46 DA+dM binaries \citep{Koester_2009}, 10 DAH \citep[magnetic DAs, see][]{Koester_2009}, 71 DB or DBA WDs \citep[with helium lines, see][]{Voss_2007}, 24 DAZ WDs (DA WDs with photospheric metal lines, generally \ion{Ca}{ii}), 25 DAs with interstellar metal lines, and 17 helium-rich WDs with metal lines \citep{Koester_2005}. As the majority of the sample are DAs, which have the sharp NLTE Balmer-line cores that permit the highest RV accuracy, we focus from here on only on a uniform sample consisting of the DA and DAZ WDs in the sample.

\citet{Falcon_2010} have measured RVs in individual SPY spectra of single DA WDs for the purpose of gravitational-redshift estimation, and we have adopted their methodology, as follows. We fitted a region of $\pm 500$\kms\ around the position of the H$\alpha$ NLTE line core with a combination of a Gaussian (fitting the NLTE core) and a parabola (fitting the local region of the full line profile), using least-squares, with RV error estimates for every fit obtained based on the covariance matrix from the fit. All velocities were barycentre-corrected using the correction values provided by the UVES pipeline. All fitted spectra were inspected by eye, and 238 epochs of 135 WDs with problematic fits due to noisy or flawed data were excluded from the sample. To optimise the statistical power of our sample, a further three spectra of three WDs having formal RV errors $> 10$\kms\ were excluded from the sample.

To examine the reliability of the RV error estimates (which is important for our Monte Carlo simulation of the results, Section~\ref{sec:monte-carlo-simul}, below), we have compared the mean RV error estimate from every pair of epochs for the same WD, to the actual RV difference between those epochs, scaled down by $\sqrt{2}$. As the majority of the WDs in the sample are single (or their binary nature is not revealed by these observations), the scaled-down $\Delta$RV values for all WDs having a given RV error estimate from the fitting process should follow a Gaussian distribution, centred on zero and with a $\sigma$ corresponding to the RV error estimate, except for some high $\Delta$RV outliers due to the minority of DWDs with real RV changes between epochs. Good agreement with this expectation is indeed reflected in Figure~\ref{fig:RVpairs}, which shows the distributions of observed RV epoch differences for several narrow ranges of the formal RV errors. From this comparison, the formal RV errors possibly underestimate the true RV errors by $\sim 0.5$\kms. Thus, we have taken 0.5\kms\ as the minimal error and updated the measured errors accordingly. The H$\beta$ NLTE line core in the spectra is not always clearly detected and gives a noisier RV measurement, and hence we rely only on H$\alpha$ for our RVs. 

In most spectroscopic DWD binaries, the light from one of the WDs is dominant, and therefore the binary is single-lined. However, 14 of the DA WDs are double-lined, with clearly separated NLTE line cores. In those cases we have fitted the region around the double NTLE core with a combination of two Gaussians and a parabola. All but a handful of the WDs have time differences between epochs of up to 470\,d, and this time difference dictates the range of orbital separations to which the sample is sensitive. We have therefore excluded from the sample epochs separated by more than 470\,d from any other epoch. Finally, the maximal velocity difference between any two epochs for every WD, \drvm, was calculated for the final sample. In the case of double-lined systems, with two RV values per epoch, the RV value used for the calculation was that of the deeper NLTE core, or chosen randomly if both NLTE cores were of the same depth.

Our final sample consists 926 spectra of 439 WDs that have more than one epoch each. Figure~\ref{fig:EpochDist} shows the distribution of the number of epochs per WD, the distribution of time differences between epochs, and the distribution of RV errors. 

\begin{figure}
\centering
\includegraphics[width=\columnwidth]{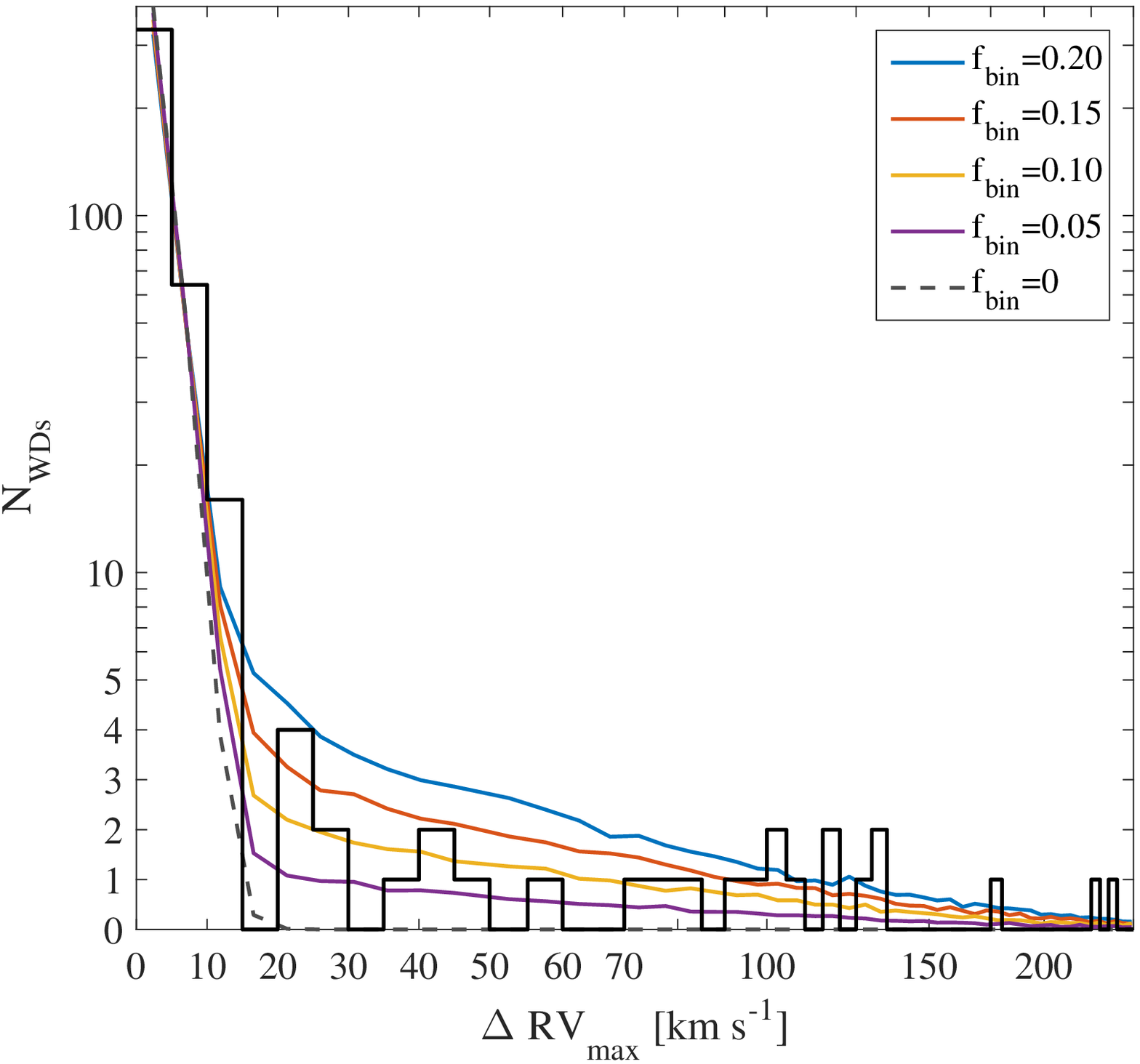}

\vspace{0.02\textheight}

\includegraphics[width=\columnwidth]{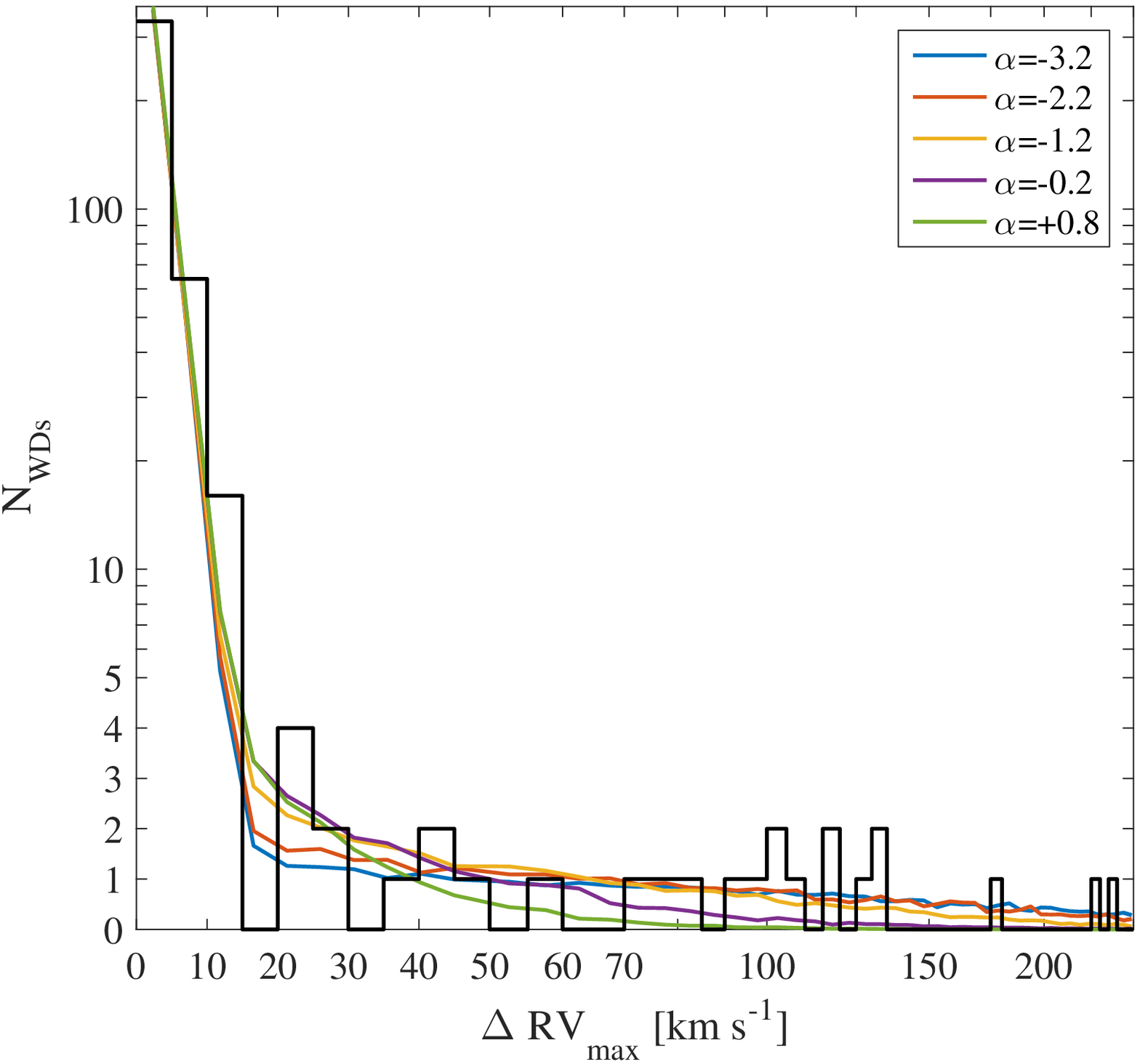}
\caption{Distribution of \drvm\ (black histogram), compared to Monte-Carlo model distributions (coloured solid curves). Top: model dependence on binary fraction, \fb, for a fixed $\alpha=-1.2$ value. The dashed black curve is a model with no DWDs (\fb$=0$). Bottom: model dependence on separation distribution power-law index, $\alpha$, for a fixed \fb$=0.10$ value.}
\label{fig:DRVM}
\end{figure}

Figure~\ref{fig:DRVM} presents the main observed statistic of this work, the distribution of \drvm, the maximal velocity difference between any two epochs for every WD. The dashed curve shows the expected \drvm\ distribution from a Monte-Carlo simulation (Section~\ref{sec:monte-carlo-simul}, below) of a sample in which there are {\it no} binaries, and therefore all observed RV changes in this simulated distribution result solely from measurement noise. One can see that, above \drvm$ \gtrsim 10$\kms, real WD binaries dominate the distribution.

Table~A\ref{tab:Full} lists the parameters and RV measurements for the full final WD sample. Table~\ref{tab:Tail} collects only the 43 candidate DWD systems having \drvm$ > 10$\kms, which are also highlighted in Table~A\ref{tab:Full}.

\begin{table*}
\caption{Candidate DWDs with \drvm > 10\kms.}
\label{tab:Tail}
\begin{center}
\begin{tabular}{l r r r r r r r l}
\hline
\multicolumn{1}{l}{Name} &
\multicolumn{1}{l}{RA} &
\multicolumn{1}{l}{Dec} &
\multicolumn{1}{c}{$\Delta t$} &
\multicolumn{1}{c}{\drvm} & 
\multicolumn{1}{c}{$\sigma$} &
\multicolumn{1}{c}{$N_\sigma$} &
\multicolumn{1}{c}{$M_1$} &
\multicolumn{1}{l}{Comments} \\
& & &
\multicolumn{1}{c}{[d]} &
\multicolumn{1}{c}{[km\,s$^{-1}$]} &
\multicolumn{1}{c}{[km\,s$^{-1}$]} &
&
\multicolumn{1}{c}{[\msun]} & \\
\hline
HE1414-0848 & 14:16:52.07 & -09:02:03.8 & 395.9 & 238.4 & 4.2 & 57.3 & 0.40 & 1, 2 \\
WD2020-425 & 20:23:59.57 & -42:24:26.7 & 23.0 & 225.9 & 2.9 & 77.3 & 0.75 & 1 \\
WD0326-273 & 03:28:48.81 & -27:19:01.7 & 3.0 & 179.3 & 1.2 & 151.2 & 0.35 & 3 \\
WD1210+140 & 12:12:33.89 & +13:46:25.1 & 0.9 & 133.1 & 1.8 & 72.6 & 0.30 & 4 \\
WD0135-052 & 01:37:59.40 & -04:59:44.9 & 4.0 & 132.4 & 0.8 & 159.3 & 0.20 & 1, 5 \\
WD0037-006 & 00:40:22.94 & -00:21:31.1 & 45.0 & 128.5 & 1.1 & 118.2 & 0.55 & 1 \\
WD0341+021 & 03:44:10.77 & +02:15:29.9 & 201.2 & 117.1 & 2.6 & 44.7 & 0.30 &  \\
WD0028-474 & 00:30:47.16 & -47:12:36.9 & 59.8 & 116.8 & 1.7 & 67.1 & 0.50 & 1 \\
HE2209-1444 & 22:12:18.05 & -14:29:48.0 & 287.1 & 106.3 & 1.4 & 75.5 & 0.60 & 1, 6 \\
WD2200-136 & 22:03:35.63 & -13:26:49.9 & 368.1 & 104.0 & 4.7 & 22.3 & 0.45 & 1 \\
WD1124-018 & 11:27:21.33 & -02:08:37.7 & 1.9 & 101.9 & 3.0 & 34.0 & 0.50 &  \\
WD1824+040 & 18:27:13.13 & +04:03:45.9 & 83.7 & 96.7 & 1.1 & 88.2 & 0.35 & 7 \\
WD0344+073 & 03:46:51.42 & +07:28:01.9 & 215.3 & 91.3 & 1.5 & 60.2 & 0.35 &  \\
WD1349+144 & 13:51:54.06 & +14:09:44.2 & 1.1 & 84.3 & 2.7 & 30.7 & 0.55 & 1, 8 \\
HS1102+0934 & 11:04:36.76 & +09:18:22.7 & 320.9 & 77.1 & 1.7 & 45.9 & 0.45 & 9 \\
HE0320-1917 & 03:22:31.91 & -19:06:47.8 & 1.0 & 70.4 & 1.2 & 61.1 & 0.30 & 10 \\
WD2330-212 & 23:32:59.48 & -20:57:12.1 & 2.9 & 55.5 & 2.6 & 21.8 & 0.40 &  \\
HE0225-1912 & 02:27:41.43 & -18:59:24.5 & 7.1 & 48.1 & 3.1 & 15.5 & 0.55 & 1 \\
WD2336-187 & 23:38:52.78 & -18:26:11.9 & 3.0 & 42.1 & 5.8 & 7.3 & 0.25 & 1 \\
HE0410-1137 & 04:12:28.99 & -11:30:08.3 & 4.0 & 40.9 & 1.4 & 28.7 & 0.50 & 1 \\
WD0032-317 & 00:34:49.82 & -31:29:54.3 & 1.0 & 38.1 & 3.8 & 10.1 & 0.35 & 11 \\
WD1013-010 & 10:16:07.01 & -01:19:18.7 & 20.9 & 29.0 & 2.3 & 12.6 & 0.25 & 12 \\
HE0325-4033 & 03:27:43.92 & -40:23:26.1 & 0.9 & 26.8 & 1.5 & 17.4 & 0.55 &  \\
HS2046+0044 & 20:48:38.26 & +00:56:00.8 & 16.9 & 23.8 & 3.8 & 6.2 & 0.70 &  \\
HE0516-1804 & 05:19:04.27 & -18:01:29.1 & 1.0 & 22.3 & 2.7 & 8.2 & 0.55 & 13 \\
HE0131+0149 & 01:34:28.46 & +02:04:21.4 & 392.9 & 21.6 & 1.3 & 16.2 & 0.50 &  \\
HE0324-1942 & 03:27:05.02 & -19:32:23.8 & 0.9 & 21.4 & 3.7 & 5.8 & 0.80 & 1 \\
HE0221-0535 & 02:23:59.88 & -05:21:45.9 & 0.9 & 14.3 & 2.0 & 7.3 & 0.60 &  \\
WD1233-164 & 12:36:14.02 & -16:41:53.5 & 270.3 & 14.1 & 3.9 & 3.7 & 0.75 &  \\
HE0221-2642 & 02:23:29.40 & -26:29:19.7 & 226.3 & 14.0 & 5.9 & 2.4 & 0.55 &  \\
HS1334+0701 & 13:36:33.67 & +06:46:26.8 & 309.2 & 13.1 & 2.0 & 6.5 & 0.40 &  \\
WD2359-324 & 00:02:32.36 & -32:11:50.7 & 2.1 & 12.6 & 3.0 & 4.2 & 0.55 &  \\
WD2253-081 & 22:55:49.49 & -07:50:03.3 & 5.0 & 12.4 & 1.7 & 7.5 & 0.20 &  \\
HS2216+1551 & 22:18:57.15 & +16:06:56.9 & 1.1 & 12.4 & 1.9 & 6.4 & 0.65 & 1 \\
HE0031-5525 & 00:33:36.03 & -55:08:37.5 & 222.2 & 12.2 & 4.6 & 2.7 & 0.45 &  \\
HE2148-3857 & 21:51:19.23 & -38:43:04.5 & 1.0 & 11.5 & 4.0 & 2.9 & 0.70 &  \\
WD2308+050 & 23:11:18.05 & +05:19:27.9 & 1.0 & 11.4 & 4.3 & 2.7 & 0.45 &  \\
WD2254+126 & 22:56:46.26 & +12:52:49.9 & 34.8 & 11.4 & 4.9 & 2.3 & 0.55 &  \\
HS1204+0159 & 12:07:29.51 & +01:42:50.6 & 1.0 & 11.3 & 4.2 & 2.7 & 0.50 &  \\
HE0344-1207 & 03:47:06.71 & -11:58:08.5 & 1.0 & 11.1 & 3.1 & 3.5 & 0.55 &  \\
WD0114-605 & 01:16:19.55 & -60:16:07.6 & 343.0 & 10.9 & 2.4 & 4.6 & 0.50 &  \\
HE0417-3033 & 04:19:22.07 & -30:26:44.0 & 261.2 & 10.2 & 3.1 & 3.3 & 0.50 &  \\
WD2248-504 & 22:51:02.02 & -50:11:31.8 & 7.0 & 10.1 & 2.8 & 3.6 & 0.60 &  \\

\hline
\end{tabular}
\end{center}

\begin{flushleft}
Notes:
$\sigma$ is the root of the summed squares of the RV errors of the two individual RV measurements forming each difference.
$N_\sigma$ is defined as \drvm$/\sigma$.
$M_1$ is the derived mass for the photometric-primary WD.

(1)~Double-lined DWD
(2)~HE1414-0848: $P=0.5178$\,d, $M_1=0.55$\,\msun, $M_2=0.71$\,\msun\ \citep{Napiwotzki_2002}
(3)~WD0326-273: $P=1.8754$\,d, $M_1=0.51$\,\msun, $M_{2,\textrm{min}}=0.59$\,\msun\ \citep{Nelemans_2005}
(4)~WD1210+140: $P=0.64194$\,d, $M_1=0.23$\,\msun, $M_{2,\textrm{min}}=0.38$\,\msun\ \citep{Nelemans_2005}
(5)~WD0135-052: $P=1.553$\,d, $M_1=0.47$\,\msun, $M_2=0.52$\,\msun\ \citep{Saffer_1988,Bergeron_1989}
(6)~HE2209-1444: $P=0.2769$\,d, $M_1=0.58$\,\msun, $M_2=0.58$\,\msun\ \citep{Karl_2003b}
(7)~WD1824+040: $P=6.26600$\,d, $M_1=0.428$\,\msun, $M_{2,\textrm{min}}=0.515$\,\msun\ \citep{MoralesRueda_2005}
(8)~WD1349+144: $P=2.2094$\,d, $M_1=0.44$\,\msun, $M_2=0.44$\,\msun\ \citep{Karl_2003a}
(9)~HS1102+0934: $P=0.55319$\,d, $M_1=0.46$\,\msun, $M_{2,\textrm{min}}=0.55$\,\msun\ \citep{Brown_2013}
(10)~HE0320-1917: $P=0.86492$\,d, $M_1=0.29$\,\msun, $M_{2,\textrm{min}}=0.35$\,\msun\ \citep{Nelemans_2005}
(11)~WD0032-317: possible 1500\,K BD companion, based on weak NIR excess.
(12)~WD1013-010: $P=0.43653$\,d, $M_1=0.44$\,\msun, $M_{2,\textrm{min}}=0.38$\,\msun\ \citep{Nelemans_2005}
(13)~HE0516-1804: possible NIR excess indicating a $T_{\rm eff}\sim 3000$\,K companion with radius $\sim 0.2$\,\rsun.
%(14)~WD1241-010: $P=3.34741$\,d, $M_1=0.31$\,\msun, $M_{2,\textrm{min}}=0.373$\,\msun\ \citep{Marsh_1995}
%(15)~WD2032+188: $P=5.0846$\,d, $M_1=0.406$\,\msun, $M_{2,\textrm{min}}=0.469$\,\msun\ \citep{MoralesRueda_2005}
\end{flushleft}
\end{table*}

A pertinent question at this point is the nature of the companions in the candidate binary systems in the tail of the \drvm\ distribution. We argue that all but one or two, at most, are also WDs (i.e. the systems are DWDs). Main-sequence companion stars that are hot enough would reveal themselves in the red regions of the SPY spectra by means of their molecular-band features. Indeed, as noted above, \citet{Koester_2009} already identified in this way 46 WD+dM systems in the SPY sample, which we have excluded from our analysis. Among the \drvm\ tail WDs, the most luminous one has absolute $R$ magnitude $M_R\sim 9.7$. An M3 companion star of effective temperature $T_{\rm eff}\approx 3300$\,K and mass $\sim 0.2$\,\msun\ has $M_R\sim 11.2$\,mag \citep{Baraffe_1996}, only a factor of 4 fainter than the WD, and hence would likely still be detected in the SPY spectrum.

To test for the presence of cooler stellar and sub-stellar companions, we have examined on VizieR \citep{Ochsenbein_2000} the available optical to near-infrared (NIR) photometry, including the spectrophotometrically calibrated SPY spectra themselves, for all of the 43 ``tail'' WDs. To the optical-band photometry of each WD, we have fit a scaled Planck spectrum with a temperature according to the WD's photospheric fit from \citet{Koester_2009}, corrected for the three-dimensional effects of convection using Tables B.5 and B.6 of \citet{Tremblay_2013}. We then searched for evidence of any NIR excess at $1-5\,\mu$m as indicated by the photometry from 2MASS \citep{Skrutskie_2006}, UKIDSS \citep{Lawrence_2007}, VISTA \citep{Jarvis_2013, Edge_2013} and WISE \citep{Cutri_2014}. To quantify any detected excesses and estimate the detection limits for excesses, we added to each model WD thermal spectrum a second thermal spectrum of varying temperature, scaled according to the ratio of the surface areas of the WD and of a $R=0.1$\,\rsun\ companion (appropriate for a low-mass star or BD). Each WD's radius was based on the effective temperature and surface gravity calculated by \citet{Koester_2009}, corrected for 3D convection \citep{Tremblay_2013}, and the theoretical WD cooling sequences of \citet{Fontaine_2001}\footnote{\url{http://www.astro.umontreal.ca/~bergeron/CoolingModels/}}. From the data, we estimate that we can detect a $\gtrsim 1500$\,K NIR excess in the spectra of the hotter WDs in the sample, and $\gtrsim 1000$\,K in the cooler WDs.
    
In four cases do we see an actual possible NIR excess. In three of these, the excess corresponds to a companion temperature of $\sim 1000-2000$\,K, i.e. a BD companion. However, two of these three cases, WD0037-006 and HE0410-1137, are double-lined DWDs, and the excess may be coming from the cooler of the two WDs in each system. Furthermore, in the double-lined spectra of these two WDs, the photometric temperature fit to the Balmer line profiles is unreliable, and this could also lead to the appearance of a too-red NIR spectral slope. In the third case, WD0032-317, a small excess seen at 3.35 and 4.6\,$\mu$m may in fact be from a BD companion. Finally, the fourth case with a NIR excess is HE0516-1804, where we find factor-2 discrepancies between different optical photometric measurements, but the NIR photometry nonetheless suggests the presence of a strong excess, corresponding to a $\sim 3000$\,K, $0.2$\,\rsun-radius emitter (such as an M4-M5 dwarf companion). The red part of the SPY spectrum is smooth, with no traces of any stellar absorption features that we would expect given such a stellar companion, but this may be expected for a companion as cool as this (see above). The nature of the companion in this case is therefore still unclear, and deserves further study. Among the 43 tail WDs, there may thus be one or two WDs whose RV variations are caused by an M star or by a $\sim 1500$\,K BD companion. 

From a statistical point of view, dedicated studies using NIR excess to discover unresolved BD companions in WD samples have estimated a WD+BD binary fraction of $0.5\pm0.3$ per cent \citep{Steele_2011}, and 0.8 to 2 per cent \citep{Girven_2011}. Thus, in the full SPY sample analysed here, one might expect a handful of WD+BD binaries at all separations out to thousands of AU. Assuming roughly equal numbers per logarithmic interval of separation, about 1/3 of those systems would be in the $<4$\,AU separation range probed by the tail WDs, i.e. one or two systems. This is consistent with our NIR-excess search results, above, for the tail WDs. In summary, the maximum contamination of WD+BD and WD+dM systems to the \drvm\ tail population is by one or two systems, and the overwhelming majority of the rest must be WDs with compact-remnant companions. Many binaries are known consisting of a WD plus a neutron star \citep[see][]{Lorimer_2001}, and systems of a WD plus a black hole also likely exist. However, such systems are expected to be $\sim 100$ times rarer than DWDs \citep[e.g.][]{Nelemans_2001} so it is unlikely that they are in our sample, let alone in significant numbers. We will therefore assume from here on that the tail of the \drvm\ distribution probes primarily the DWD population.
  
As an aside, among the 19 DAZ WDs with a photospheric \ion{Ca}{ii}\,K ($\lambda$3933.7\,\AA) absorption line \citep{Koester_2005} that have multiple epochs, we tested for any changes in the equivalent width of the lines between epochs. The time differences span the full range of the sample -- some are a few days, some a few weeks, and some several months to a year. Photospheric metal absorption in WDs results from recent or ongoing accretion of planetary debris \citep[e.g.][]{Jura_2003, Koester_2014}, and it is not unreasonable to expect that the WD photospheric abundances could change on these timescales due to changes in accretion rate and gravitational settling of the accreted ions. However, we detect no changes in \ion{Ca}{ii}\,K equivalent width in any of these WDs, to $\lesssim 0.03$\,\AA.

\section{Monte Carlo simulation of the \drvm\ distribution of the binary WD population observed by SPY}
\label{sec:monte-carlo-simul}

To use the observed \drvm\ distribution to set constraints on the DWD population, we now simulate families of assumed binary WD populations. We then ``observe'' each simulated population with the sampling sequences and the velocity error distributions of the real data, to produce a model \drvm\ distribution for each simulated population. Except for a number of minor updates, our methodology follows closely the one in \citetalias{Maoz_2012} and \citetalias{Badenes_2012}. For convenience, we re-describe it here briefly.  

As noted in \citetalias{Maoz_2012}, our modelling approach is distinct from that of ``binary population synthesis'' (BPS) calculations \citet[e.g.][]{Ruiter_2009, Mennekens_2010, Toonen_2011}. In BPS, one simulates the evolution of a population of main-sequence binaries, from its initial mass and separation distributions, through the various stages of stellar and binary evolution, including mass transfer, mass loss, and common envelope. Due to the multiplicity of free parameters for the initial conditions, and the highly simplified treatment of the physical processes, there is large variety in the predictions of different BPS models. Here, in contrast, we parametrise the properties of the DWD population only at end of last common-envelope phase, after which gravitational wave emission alone drives the decay of the orbit. By skipping over the previous phases of evolution, our approach permits testing regions of the parameter space describing the true DWD population that may not be accessible to BPS, because no calculated BPS models have led to those regions.

To simulate a WD system (single or binary) we first assign it a primary mass, $m_1$ (`primary' and `secondary' refers here to the larger and smaller mass, respectively, not the photometric property). The primary mass is drawn from the observed WD mass function found by \citet{Kepler_2015} for $1504$ hot (effective temperature $T_{\rm eff} > 12000$\,K) DA WDs with $S/N>10$ in the DR10 SDSS catalogue. We use the \citet{Kepler_2015} representation of the WD mass distribution with three Gaussian components -- a main, narrow, component centred at 0.65\,\msun, with $1\sigma$ width of 0.044\,\msun, a second component centred at 0.57\,\msun, of width 0.097\,\msun, and height 0.17 of the main component, and a third component centred at 0.81\,\msun, of width 0.187\,\msun, and height 0.06 of the main component. Compared with the WD mass function of \cite{Kepler_2007}, used in \citetalias{Maoz_2012} and \citetalias{Badenes_2012}, this updated WD mass function is mildly changed: its peak is at slightly higher masses, by about 0.05\,\msun, it has a more prominent component of $\gtrsim 0.9$\,\msun WDs, and it has a less prominent component of $\lesssim 0.4$\,\msun WDs. We find that only the low-mass component in the \cite{Kepler_2007} mass distribution changes the calculated model \drvm\ distributions and, if used, it mainly shifts the allowed range of model values of the binarity fraction \fb, up by $\sim 0.02$. However, the primaries in DWDs likely have a mass distribution distinct from that of single WDs, particularly in close DWDs that have been affected by mass transfer and mass loss. We further study in Section~\ref{sec:Results} the effect of the assumed primary mass distribution on our conclusions.

Our final sample's RV precision and the time intervals between epochs bound the range of WD separations within which RV changes can indicate binarity and hence within which the binary fraction \fb\ can be constrained. In an extreme-mass-ratio WD binary with masses $m_1=1.2$\,\msun\ and $m_2=0.2$\,\msun\ and separation $a$, the secondary ($m_2$) will have circular velocity and period
\begin{equation}
v_{\rm circ}=30\,{\rm km\,s}^{-1} (a/{\rm AU})^{-1/2}, ~~~ P=308\,{\rm d}~
(a/{\rm AU})^{3/2},
\end{equation}
respectively. If the system is optimally inclined to our line of sight ($i=90^\circ$), and it is observed at the two quadrature phases, i.e. with an epoch separation $\Delta t=P/2$, yielding an RV velocity difference between epochs of $\Delta$RV$=2v_{\rm circ}$, then
\begin{equation}
\frac{\Delta{\rm RV}}{\Delta t}=\frac{4\times 30\,{\rm km\,
    s}^{-1}}{308\,{\rm d}}\left(\frac{a}{{\rm AU}}\right)^{-2}.   
\end{equation}
Thus, the largest-separation DWD systems to which the sample is sensitive have
\begin{equation}
a=1~{\rm AU}\left(\frac{\Delta{\rm RV}}{120\,{\rm km\,
    s}^{-1}}\right)^{-1/2} \left(\frac{\Delta t}{308\,{\rm d}}\right)^{1/2}.
\end{equation}
Our sample, with a minimum detectable $\Delta{\rm RV}\approx 10$\kms\ and maximum separation between epochs of $\Delta t=470$\,d, is thus sensitive to DWD systems with separations out to $a\approx 4$\,AU. We therefore define \fb\ as the fraction of all WD systems (both single systems and binary systems) that are binary systems with separations $a<4$\,AU.

WDs with masses less than $m_{\rm lim}\approx 0.45$\,\msun\ cannot form in isolation over a Hubble time, and therefore must have been in close-separation binaries such that interactions affected the stellar evolution of the WD progenitors. Indeed, ELM WDs with masses $<0.25$\,\msun\ are always observed to be in binaries \citep{Brown_2016}. Nonetheless, 30 per cent of WDs with masses 0.32\,\msun$<M<$0.45\,\msun\ may be single after all \citep{BrownJ_2011}, with a number of mechanisms having been proposed for this singularity \citep[see][]{BrownJ_2011, Zorotovic_2016}. We account for these observations by assigning binary companions to 70\% of the simulated WD primaries in the $0.25-0.45$\,\msun\ mass range. (In this we differ from the treatment in \citetalias{Badenes_2012}, where WDs in this mass range had a probability \fb\ for binarity.) Simulated primary WDs with mass $<m_{\rm lim}=0.25$\,\msun, are always assigned to binaries. The fraction of the \cite{Kepler_2015} mass function that is either below $m_{\rm lim}=0.25$\,\msun, 0.025\%, or between 0.25\,\msun\ and 0.45\,\msun, 2.7\%, means that 0.025\% of the simulated WDs in the sample are in binaries with one of the WDs having $<0.25$\,\msun, $0.7\times 2.7\%=1.9\%$ are binaries with one WD in the $0.25-0.45$\,\msun\ range, and a fraction \fb$-\left(0.00025+0.7\times 0.027\right)$ of simulated WDs are in binary systems in which both components have masses above $m_{\rm lim}=0.45$\,\msun.

To the simulated WDs chosen to be in binaries, additional binary parameters are assigned, as we will describe below. To the remaining $1-$\fb\ fraction of the WDs are given an orbital velocity of zero, and the simulation goes directly (see below) to the allocation of random velocity errors at several observing epochs.

Following \citetalias{Maoz_2012}, we draw the secondary WD mass, $m_2$, from a power-law distribution in mass ratio,
\begin{equation}
P(q)\propto q^\beta ,~~~~
q\equiv m_2/m_1 ,
\end{equation}
with $m_2$ between 0.45\,\msun\ and $m_1$. The parameter $\beta$ can be zero (equal probability for all mass ratios in the range), positive (preference for similar-mass DWDs), or negative (preference for low-mass companions). The power-law index $\beta$ is one of the parameters of the WD binary population that could, in principle, be constrained by the data, but in practice we find that the \drvm\ distribution is only weakly sensitive to it (as shown also in \citetalias{Badenes_2012} for the SDSS WD dataset). In cases where the primary in the Monte-Carlo draw was below 0.25\,\msun\ (and hence the WD is always in a binary), or the primary is in the 0.25\,\msun\ to 0.45\,\msun\ range (and hence it has a 70\% probability of being in a binary), then $m_2$ is chosen with equal probability between $0.2$\,\msun\ and $1.2$\,\msun.

We assume in our simulations an initial power-law WD separation distribution at the end of the final common-envelope phase, with an index $\alpha$ that is a to be constrained by the observations. Apart from the physical and observational arguments for such a functional form (see \citetalias{Maoz_2012} and \citetalias{Badenes_2012}), the range of separations that we consider here, $\sim 0.001-4$\,AU, is limited enough that a power-law can approximate a broad range of other, monotonic, functional dependences. Orbital decay due to gravitational wave emission will modify any initial separation distribution with time, as all of the orbits shrink, and the tightest systems merge. The distribution at a given time will be the integral over populations of different ages. \citetalias{Maoz_2012} calculated analytic expressions for this evolved, time-integrated, distribution of DWD separations, under some simplifying assumptions, and we briefly repeat here the salient points.

The separation $a$ between two point masses, $m_1$ and $m_2$, in a circular orbit shrinks over time due to gravitational-wave losses as
\begin{equation}
\frac{da}{dt}=-\frac{K}{4 a^{3}},~~~~
K\equiv\frac{256}{5}\frac{G^3}{c^5}m_1 m_2 (m_1+m_2).  
\end{equation}
The time $t$ to evolve from separation $a'$ to separation $a$ follows
\begin{equation}
\label{timetomerger}
a'^4-a^4=Kt.
\end{equation}
Assume that a co-eval population of WD binaries forms at a time $t=0$, after having emerged from their final common envelope phase, with an initial distribution of separations $n'(a')$, assumed independent of separation. For the evolved distribution $n(a,t)$, conservation of the number of systems,
\begin{equation}
n(a,t) da=n'(a') da' ,
\end{equation}
and assuming the initial distribution is a power law,
\begin{equation}
n'(a')\propto a'^\alpha,
\end{equation}
then gives
\begin{equation}
n(a,t)\propto a^3 (a^4+Kt)^{(\alpha-3)/4},
\label{eqsinglepop}
\end{equation}
a broken power law. At separations $a\gg(Kt)^{1/4}$ (i.e., much larger than those that can merge within time $t$), the distribution has the original power-law slope, $n(a)\sim a^\alpha$. For $a\ll(Kt)^{1/4}$, $n(a)\sim a^3$ (see figure~3 of \citetalias{Maoz_2012}).

Assuming a series of binary WD generations, produced at a constant rate between $t=0$ and the current age of the Galaxy, $t_0$, the present separation distribution is
\begin{equation}
N(x)\propto x^{4+\alpha} [(1+x^{-4})^{(\alpha+1)/4}-1],~~~~~ \alpha\ne -1,
\label{eqtimeint}
\end{equation}
or
\begin{equation}
N(x)\propto x^{3} \ln(1+x^{-4}),~~~~~ \alpha= -1,
\label{eqtimeintlog}
\end{equation}
where
\begin{equation}
x\equiv\frac{a}{(K t_0)^{1/4}}
\end{equation}
is the separation scaled to that of binaries that merge within the Galaxy lifetime. For example, for $t_0=13.6$\,Gyr and $m_1=m_2=0.6$\,\msun, $x=1$ corresponds to $a_0\approx 2.5$\,\rsun. $N(x)$ is, again, roughly a broken power law, with index $\alpha$ at $x \gg 1$. At $x \ll 1$ the power-law index is 3 for $\alpha\ge -1$, and $\alpha + 4$ for $\alpha\le -1$. Even if the star-formation history is ``bumpy'' rather than constant, as assumed above, the WD formation history will be the convolution of the star-formation history with a broad, $\sim t^{-0.5}$, kernel, that describes the WD supply rate from a coeval single stellar population, which will smooth out the WD production rate. Therefore Eqns.~\ref{eqtimeint}-\ref{eqtimeintlog} hold, as long as the star-formation history falls less steeply than $\sim t^{-0.5}$.

In every realization of our simulations, we draw each simulated DWD separation from the distributions described by Eqns.~\ref{eqtimeint}-\ref{eqtimeintlog} for a particular value of $\alpha$, with $a$ between $a_{\rm min} =2\times 10^4$\,km (DWD contact) and $a_{\rm max}=4$\,AU. For each simulated DWD system, Kepler's law gives the period
\begin{equation}
\tau=2\pi \left(\frac{a^3}{G(m_1+m_2)}\right)^{1/2},
\end{equation}
and the (assumed circular) orbital velocities,
\begin{equation}
v_1=\frac{2\pi a}{\tau}\frac{m_2}{m_1+m_2}, ~~~v_2=\frac{2\pi
  a}{\tau}\frac{m_1}{m_1+m_2} .
\end{equation}
Binaries with periods $<10$\,min are assigned zero orbital velocity, as the exposure times of the individual VLT spectra prevent detection of velocity differences in such cases. 

The merger rate per WD, for a given set of parameters, is obtained numerically by counting the fraction of all of the systems in the simulation whose merger lifetime, $t_{\rm merge}=a^4/K$ (Eq.~\ref{timetomerger} with $a=0$) is smaller than some time interval, divided by that time interval. Recent observational estimates of the Milky Way's disk space density of WDs are all broadly consistent and include $0.0046\pm0.0005$\,pc$^{-3}$, from a proper-motion selected sample of WDs in SDSS \citep{Harris_2006}, $0.0049\pm 0.0005$\,pc$^{-3}$ for a sample within 20\,pc \citep{Sion_2009}, $0.0048\pm 0.0005$\,pc$^{-3}$ for a sample within 25\,pc \citep{Holberg_2016}, and $0.0055\pm 0.0001$\,pc$^{-3}$ using a large new proper motion catalogue of SDSS WDs \citep{Munn_2016}. The local stellar mass density is 0.085\,\msun\,pc$^{-3}$ \citep{McMillan_2011}. Dividing it by the WD space density of \citet{Munn_2016}, there are 15.5\,\msun\ of stellar mass in the disk per WD. We therefore divide the DWD merger rate per WD by 15.5\,\msun\ to convert it to a DWD merger rate per unit stellar mass. Adopting the slightly lower value of \citet{Holberg_2016} for the WD density would systematically lower all of our model merger rates by 15 per cent.

A line-of-sight inclination $i$ angle of the the orbital plane is chosen for each system and the line-of-sight velocity is reduced by $\sin i$. The photometric primary WD, which determines which of the two WD's RVs is measured, can be either the less or more massive WD, depending on the WD surface areas (dictated mainly by mass) and temperatures (set by cooling age, and potentially complicated by interactions between the binary components). Following the observational and theoretical considerations in \citetalias{Maoz_2012}, we make the less massive WD the photometric primary when its mass is below 0.35\,\msun, but decide randomly, with equal probability, between the two WDs when the less massive WD is above this limit. We have tested the sensitivity of the results to this assumption by trying {\it always} to take the lower-mass WD as the photometric primary. We find negligible changes.

The line-of-sight velocity (i.e. RV) curve of the photometric primary is sampled with the actual distribution of temporal samplings in our SPY sample, by applying at random a particular real observation pattern (number of epochs and time between epochs) from the sample. Every simulated velocity measurement is noised with a random error, drawn from a Gaussian distribution. The variance of the Gaussian, in turn, is drawn from the distribution of measurement errors of the observed sample (see Figure~\ref{fig:EpochDist}). Finally, to calculate \drvm, we find the difference between the minimum and maximum observed velocities for every simulated DWD or single WD. We create $4\times 10^5$ WD systems (some single, some binary, according to \fb) for every parameter combination that defines a DWD population model. The fractional prediction for each bin in the model \drvm\ distribution, multiplied by the observed WD sample size, gives the expectation value for that bin.

\citetalias{Maoz_2012} discussed how, for the SDSS sample, the \drvm\ distribution depends on the binary population parameters \fb, $\alpha$, and $\beta$. We show this in Figure~\ref{fig:DRVM} for the SPY sample, with its much higher RV resolution, which probes DWDs at larger separations and smaller orbital velocities. The \drvm\ distribution strongly discriminates in \fb, which affects the amplitude in the $\sim 10-100$\kms\ range. Changes in $\alpha$ affect the slope of the distribution in the range, but a larger relative range is consistent with the observations, and only more extreme values of $\alpha$ are excluded. As already noted, the \drvm\ distribution depends weakly on $\beta$, the power-law index of the binary mass-ratio distribution, so RV survey data of the type considered here do not constrain this binary population characteristic. Conversely, not knowing the distribution of mass ratios does not affect adversely our ability to constrain the other binary population parameters.
 
To compare each simulated model \drvm\ distribution to the observed one, we take the model expectation value for each velocity bin in the \drvm\ distribution, and we sum the logarithms of the Poisson probabilities of finding the observed number of systems in each bin, given the expectations from the model. This gives the log of the likelihood of each model. We run models over a grid in parameter space, to find the allowed region of the DWD population parameter space.

\section{Results}
\label{sec:Results}

\begin{figure}
\centering
\includegraphics[width=\columnwidth]{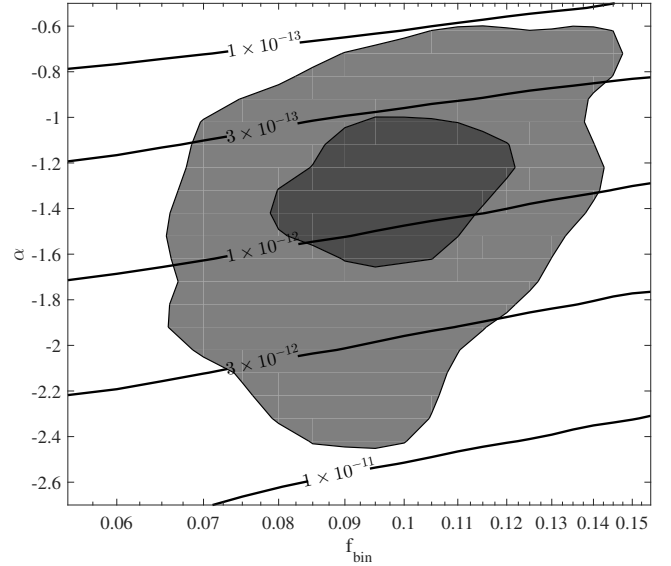}
\caption{Likelihood contours in the \fb, $\alpha$ plane, indicating the $1\sigma$ (dark gray) and $2\sigma$ (light gray) confidence levels. The overlaid lines are curves of constant WD merger rate in ${\rm yr}^{-1}$\,\msun$^{-1}$, as marked. The Milky Way's specific SN Ia rate is $R_{\rm Ia}\approx 1.1\times 10^{-13}$\,yr$^{-1}$\,\msun$^{-1}$}
\label{fig:Contours}
\end{figure}

Figure~\ref{fig:Contours} shows the parameter space of \fb\ and $\alpha$, with contours showing the $1\sigma$ and $2\sigma$ likelihood ranges, corresponding to changes of 0.5 and 2, respectively, in log-likelihood compared to the best-fit model. Models acceptable at the $2\sigma$ level occupy a well-defined region in the \fb$-\alpha$ plane, with \fb\ going from $\sim$0.07 to 0.14 as $\alpha$ goes from -2.4 to -0.6. The best-fit $1\sigma$ ranges are \fb$= 0.100 \pm 0.020$ and $\alpha=-1.32 \pm 0.30$. These constraints are significantly improved compared to those from the SDSS WD sample in \citetalias{Badenes_2012}. Also shown in Figure~\ref{fig:Contours} are curves of constant merger rate, which in this kind of presentation (linear in $\alpha$, logarithmic in \fb) appear roughly as straight lines (see \citetalias{Maoz_2012}). The acceptable models (over the $2\sigma$ region) span merger rates of $R_{\rm merge}=1\times 10^{-13}\,{\rm yr}^{-1}$\,\msun$^{-1}$ to $8\times 10^{-12}\,{\rm yr}^{-1}$\,\msun$^{-1}$. The likelihood-weighted merger rate over the $1\sigma$ region is $(7 \pm 2)\times 10^{-13}\,{\rm yr}^{-1}$\,\msun$^{-1}$.

The observed \drvm\ distribution (Figure~\ref{fig:DRVM}) identifies 43 systems that are candidate DWDs in the distribution's ``tail'' (i.e. beyond the ``core'' of the distribution that is produced by RV errors in single WDs, and by the DWDs that are not fortuitously time-sampled or insufficiently inclined to the line of sight). See Table~\ref{tab:Tail} for the full candidate list. 27 of them are very likely DWD systems (with \drvm>15\kms) and a further 16 (with 10\kms<\drvm < 15\kms) are possible DWDs. Among the 43, 13 cases are double-lined DWDs. \citet{Koester_2009} have derived and compiled $T_{\rm eff}$ and log surface gravity ($\log g$) estimates for the SPY WDs from modelling of the WD absorption line profiles with synthetically calculated WD atmosphere spectra. They note that the atmospheric fits assumed a single WD, and hence the results for the double-lined systems are approximate. As described in Section~\ref{sec:Sample}, we corrected for 3D effects using \citet{Tremblay_2013}, and used the theoretical WD cooling sequences of \citet{Fontaine_2001} to derive masses for all of the WDs. We list them in Tables~\ref{tab:Tail} and A\ref{tab:Full}. 

\begin{figure}
\centering
\includegraphics[width=\columnwidth]{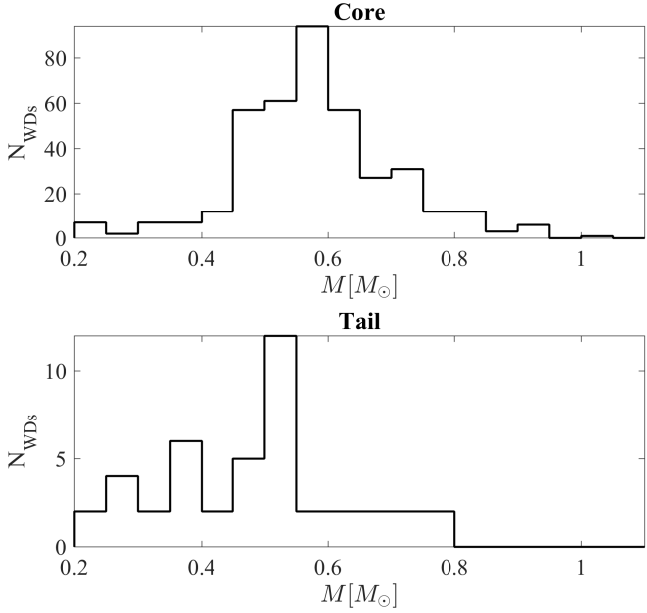}
\caption{Estimated mass distributions of the mostly single WDs from the ``core'' of the \drvm\ distribution (top panel) and of the visible photometric primaries among the DWDs from the \drvm\ distribution's $>10$\kms\ ``tail'' (bottom panel).}
\label{fig:Mass}
\end{figure}

Figure~\ref{fig:Mass} compares the mass distributions of the photometric-primary WDs in our full SPY sample and for the likely DWD systems (from the ``tail'' of the \drvm\ distribution). The mass distribution of the SPY WDs is quite similar to that of \cite{Kepler_2007} for SDSS WDs in the sense of having a significant low-mass WD component. The photometric primaries among the SPY DWD candidates, however, are clearly biased to lower masses. This is not surprising, since most or all of these systems must have undergone common-envelope evolution, which can stunt the growth of the degenerate core of the evolved star in the system, and thus produce WDs below the $\sim 0.45$\,\msun\ Hubble-time stellar-evolution limit.

The question naturally arises as to the masses of the unseen photometric secondary WDs in the DWD systems. Unfortunately, the SPY few-epoch observations cannot answer this question. More intensive RV measurements exist for 14 of these systems \citep{Saffer_1988,Bergeron_1989,Marsh_1995,Napiwotzki_2002,Karl_2003a,Karl_2003b,Nelemans_2005,MoralesRueda_2005,Brown_2013}. There is some indication that the companion masses, also listed for these cases in Tables~\ref{tab:Tail} and A\ref{tab:Full}, tend to be similar to, or slightly larger than, those of the photo-primary WDs, but the number is still too small for a clear picture of the photo-secondary mass distribution. However, it is interesting to note that for the sample of 62 ELM $\sim 0.2$\,\msun\ WDs of \cite{Brown_2016}, the unseen photo-secondary WDs have a broad mass distribution with a mean mass about four times higher than the ELMs, with a significant fraction of companion WDs at masses $>0.9$\,\msun. It is at-least conceivable that a similar trend exists for the DWD systems in SPY, with the relatively low-mass photometric-primary WDs accompanied, in many cases, by $> 0.9$\,\msun\ or $> 1.0$\,\msun\ WDs. Recent ``violent merger'' hydrodynamical simulations of DWD mergers \citep{Pakmor_2012,Ruiter_2013,Pakmor_2013} suggest that if one of the merging WDs is above such a mass, an off-center ignition can be set off in the accretion flow onto that WD, and can in turn set off a detonation wave through its high-density interior, producing an explosion that agrees with observed SN Ia properties. If the true DWD merger rate is sufficiently in the high end of the allowed range in Figure~\ref{fig:Contours}, Section~\ref{sec:monte-carlo-simul}, when we experimented with replacing the \citet{Kepler_2015} plus a large-enough fraction of DWDs have a massive WD binary component, and the violent-merger mechanism works, then some or even all SN Ia events would be explainable as DWD mergers.

To test whether the type of mass distributions for the DWD components just discussed are still consistent with the observed SPY \drvm\ distribution and the allowed region of \fb-$\alpha$ parameter space that we have delineated, we have run several more simulation grids of models. Instead of choosing the mass-primary WD from the \citet{Kepler_2015} WD mass function, we chose the mass primary from a three-Gaussian fit to the actual mass distribution of the SPY photo-primaries in Figure~\ref{fig:Mass} (top panel). The Gaussian components are a main component centred at 0.55\,\msun, with $1\sigma$ width of 0.07\,\msun, a second component centred at 0.72\,\msun, of width 0.1\,\msun, and height 0.24 of the main component, and a third component centred at 0.3\,\msun, of width 0.078\,\msun, and height 0.10 of the main component. The secondary mass selection and the rest of the simulation procedure remain as before. We find that, with this SPY-sample mass distribution, the allowed regions in \fb-$\alpha$ parameter space change slightly: the \fb\ values of the allowed regions are shifted up by $\approx 0.02$, while 1$\sigma$ region expands to include also $\alpha$ values that are more negative by $\sim -0.2$. (This trend is not surprising since, as noted above, the SPY WD mass distribution somewhat resembles that of \citet{Kepler_2007}, and as described in Section~\ref{sec:monte-carlo-simul}, when we experimented with replacing the \citet{Kepler_2015} mass function with the \citet{Kepler_2007} mass function, we obtained a similar shift.)

As a further test of the systematic dependence of our results on the assumed simulated DWD masses, we have repeated the experiment, but now drawing the {\it photo}-primary of every simulated DWD from a 3-Gaussian fit to the observed mass distribution of the 43 DWD candidates from the tail of the \drvm\ distribution (Figure~\ref{fig:Mass}, bottom panel). Here, the Gaussian components are a main component centred at 0.51\,\msun, with $1\sigma$ width of 0.024\,\msun, a second component centred at 0.35\,\msun, of width 0.087\,\msun, and height 0.35 of the main component, and a third component centred at 0.68\,\msun, of width 0.1\,\msun, and height 0.18 of the main component. The unseen second WD was chosen from a broad Gaussian centred at 0.75\,\msun, with $\sigma=0.25$\,\msun, similar to the one deduced by \citet{Brown_2016} for the mass distribution of the ELM WD companions. Compared to the results in Figure~\ref{fig:Contours}, Section~\ref{sec:monte-carlo-simul}, where we assumed the \citet{Kepler_2015} mass fuction, with these choices the allowed contours shift vertically up in $\alpha$ by $\sim 0.2$. Considering the model uncertainty in the distributions of the DWD masses, as explored by this range of experiments, we adopt {\it systematic} uncertainties $+0.02$ in \fb\ and $\pm 0.2$ in $\alpha$, compounding the random uncertainties, discussed previously.

\section{Conclusions}
We have measured and analysed the distribution of maximum radial velocity differences between observing epochs, \drvm, for a sample of 439 DA-type WDs from the SPY program, and have modelled the \drvm\ distribution to set constraints on the properties of the DWD population. Assuming that every generation of the DWD population, when it emerges from its last common-envelope phase, has an initial separation distribution that can be represented by a power law over the $a<4$\,AU separation range probed by these data, then the fraction of all WDs that have companion WDs in this separation range is \fb=$0.10 \pm 0.020 (1\sigma) +0.02$ (systematic), and the power-law index of the separation distribution is $\alpha=-1.3 \pm 0.30 (1\sigma)\pm 0.2$ (systematic). Combined with current estimates of the local WD space density and the local stellar mass density, these parameters imply a gravitational-wave-loss-driven specific Milky Way WD merger rate of $1\times 10^{-13}$ to $8\times 10^{-12}\,{\rm yr}^{-1}$\,\msun$^{-1}$ ($2\sigma$ range). This is between 1 to 70 times the estimated Milky-Way SN~Ia rate per unit mass. If some fraction (perhaps as small as a few per cent) of DWD mergers can produce a normal SN~Ia explosion, our results imply that there is no shortage of the progenitor DWD population for this explosion scenario.

Our results indicate about a 3\% DWD binary fraction per decade in separation among WDs. \citet{Klein_2017} have recently estimated the binary fraction among DWD stellar progenitors, where both progenitors have $>1$\,\msun, and found a 4\% fraction per decade in period, corresponding to 6\% per decade in separation, with some overlap between the separation ranges probed by their analysis and ours. In a Galactic population of WDs in binaries, for roughly half of the systems with WD companions having initial mass $>1$\,\msun, the companions will not yet have evolved into a WD, and therefore the companions will completely dominate the light, preventing the inclusion of such systems in WD samples. The 3\% fraction of DWDs per decade in separation that we see in SPY is therefore nicely consistent with the 6\% per decade found by \citet{Klein_2017} for DWD progenitors.

The high-\drvm\ tail of the distribution identifies 27 very likely DWD systems (with \drvm>15\kms; in two cases the WD companions may be a BD and a cool M star) and a further 16 possible DWD systems (with 10\kms< \drvm < 15\kms). The \drvm\ distribution does not constrain the component masses of the DWDs, and thus these DWDs merit follow-up observations to confirm their nature and to derive their orbital parameters and component masses. From photospheric modelling, the masses of the photometric primaries in the DWDs tend to be of somewhat lower mass than typical single WDs, with a broad distribution centred around $\sim 0.5$\,\msun. The minority of these DWDs that have additional data in the literature generally have, for the unseen photo-secondary WDs, masses similar to, or somewhat larger than, the photo-primary.

The ``double-degenerate'' scenario for SNe Ia, invoking DWD mergers as the progenitors of SNe~Ia, has been traditionally criticised on two main grounds \citep[see][]{Maoz_2014}. The numbers and hence the merger rate of the progenitor DWD populations was thought to be too small to match the SN~Ia rate, particularly if a total merged mass above the Chandrasekhar mass is required for an explosion, as often assumed in this scenario. The second long-standing problem has been theoretical -- the tidal disruption of the secondary-mass WD by the primary, and its gradual accretion onto the primary WD through a disk or a spherical configuration was thought to lead to either a stable, just more-massive, merged WD or, in the case of an above-Chandrasekhar final mass, to an ``accretion-induced collapse'' to a neutron star and an electron-capture supernova explosion. (However, the accretion-induced collapse outcome has emerged from one-dimensional calculations, which could change in 3D treatments, that are yet to be performed.)
 
Our results for the DWD population are germane to both of these objections to the double-degenerate SN~Ia scenario. It is possible that follow-up observations of the SPY DWDs will reveal, in analogy to what has been found for ELM WDs, that their WD companions have a broad mass distribution, with a significant fraction at masses above 0.9 or 1\,\msun. If so, and given that our results allow for up to a factor-70 surplus in the total rate of DWD mergers, there could conceivably be a sufficient number of mergers of CO+CO WDs with above-Chandrasekhar merged masses. Alternatively, recent violent-merger models \citep[e.g.][]{Pakmor_2012} find that an above-Chandrasekhar total mass is not required for reproducing a normal SN~Ia explosion, but rather only a primary mass above 0.9 or 1\,\msun. The secondary, which could even be a low-mass He WD, serves only as a ``hammer'' that sets off the detonation in the primary. If this were true, then an even-larger fraction of the total DWD merger rate could lead to a SN~Ia explosion. A remaining problem for violent mergers is the asymmetry of the explosion predicted by current models, and manifested in the expected strong (but unobserved in practice) polarisation of the light from SNe~Ia \citep{Bulla_2016}.

On the observational side, progress is achievable from follow-up observations and detailed characterisation of the individual DWD systems. Upcoming large and complete WD and DWD samples from the {\it Gaia} Mission will bring into much better focus the DWD population and its merger rate. In parallel, continued improvements in the theoretical study of WD mergers should further clarify if and under what conditions mergers can lead to SN~Ia explosions.

\section*{Acknowledgements}
We thank the original SPY survey team for their efforts in proposing and obtaining this valuable observational dataset.
We are grateful to Carles Badenes, Ira Bar, Uri Malamud (who suggested the \ion{Ca}{ii}\,K equivalent width test), and Yossi Shvartzvald for their contributions to this work.
The anonymous referee is thanked for constructive criticisms that improved this paper.
This work was supported by Grant 648/12 by the Israel Science Foundation (ISF) and by Grant 1829/12 of the I-CORE program of the PBC and the ISF. We acknowledge the hospitality of the Munich Institute for Astro- and Particle Physics, where this work was completed.
Based on data obtained from the ESO Science Archive Facility for programmes 165.H-0588 and 167.D-0407.
This research has made use of the VizieR catalogue access tool, CDS, Strasbourg, France.
This publication makes use of data products from the Two Micron All Sky Survey, which is a joint project of the University of Massachusetts and the Infrared Processing and Analysis Center/California Institute of Technology, funded by the National Aeronautics and Space Administration and the National Science Foundation.
This publication makes use of data products from the Wide-field Infrared Survey Explorer, which is a joint project of the University of California, Los Angeles, and the Jet Propulsion Laboratory/California Institute of Technology, funded by the National Aeronautics and Space Administration.

%%%%%%%%%%%%%%%%%%%%%%%%%%%%%%%%%%%%%%%%%%%%%%%%%%



\bibliographystyle{mnras}
\bibliography{SPY} % if your bibtex file is called example.bib

%\begin{thebibliography}{99}
%\end{thebibliography}

%%%%%%%%%%%%%%%%%%%%%%%%%%%%%%%%%%%%%%%%%%%%%%%%%%

%%%%%%%%%%%%%%%%% APPENDICES %%%%%%%%%%%%%%%%%%%%%

\appendix

\begin{table*}
\caption{WD sample.}
\label{tab:Full}
\begin{center}
\begin{tabular}{l r r r r r r r l}
\hline
\multicolumn{1}{l}{Name} &
\multicolumn{1}{l}{RA} &
\multicolumn{1}{l}{Dec} &
\multicolumn{1}{c}{$\Delta t$} &
\multicolumn{1}{c}{\drvm} & 
\multicolumn{1}{c}{$\sigma$} &
\multicolumn{1}{c}{$N_\sigma$} &
\multicolumn{1}{c}{$M_1$} &
\multicolumn{1}{l}{Comments} \\
& & &
\multicolumn{1}{c}{[d]} &
\multicolumn{1}{c}{[km\,s$^{-1}$]} &
\multicolumn{1}{c}{[km\,s$^{-1}$]} &
&
\multicolumn{1}{c}{[\msun]} & \\
\hline
{\bf WD2359-324} & {\bf 00:02:32.36} & {\bf -32:11:50.7} & {\bf 2.1} & {\bf 12.6} & {\bf 3.0} & {\bf 4.2} & {\bf 0.55} & {\bf } \\
WD0000-186 & 00:03:11.21 & -18:21:57.6 & 0.9 & 0.2 & 1.5 & 0.2 & 0.50 &  \\
WD0011+000 & 00:13:39.19 & +00:19:23.1 & 3.0 & 3.1 & 1.2 & 2.6 & 0.50 &  \\
WD0013-241 & 00:16:12.63 & -23:50:06.4 & 1.0 & 2.6 & 1.2 & 2.2 & 0.60 &  \\
WD0016-258 & 00:18:44.49 & -25:36:42.2 & 1.0 & 3.5 & 1.7 & 2.1 & 0.50 &  \\
WD0016-220 & 00:19:28.23 & -21:49:04.9 & 0.9 & 1.6 & 0.9 & 1.7 & 0.55 &  \\
WD0017+061 & 00:19:40.99 & +06:24:06.2 & 8.1 & 2.6 & 2.9 & 0.9 & 0.50 &  \\
WD0018-339 & 00:21:12.90 & -33:42:27.4 & 3.0 & 1.7 & 1.1 & 1.5 & 0.75 &  \\
WD0024-556 & 00:26:41.08 & -55:24:44.9 & 2.0 & 3.4 & 1.5 & 2.2 & 0.85 &  \\
{\bf WD0028-474} & {\bf 00:30:47.16} & {\bf -47:12:36.9} & {\bf 59.8} & {\bf 116.8} & {\bf 1.7} & {\bf 67.1} & {\bf 0.50} & {\bf 1} \\
WD0029-181 & 00:32:30.33 & -17:53:23.3 & 0.9 & 1.5 & 1.5 & 1.0 & 0.65 &  \\
{\bf HE0031-5525} & {\bf 00:33:36.03} & {\bf -55:08:37.5} & {\bf 222.2} & {\bf 12.2} & {\bf 4.6} & {\bf 2.7} & {\bf 0.45} & {\bf } \\
HE0032-2744 & 00:34:37.91 & -27:28:20.0 & 3.0 & 5.5 & 2.6 & 2.1 & 0.55 &  \\
{\bf WD0032-317} & {\bf 00:34:49.82} & {\bf -31:29:54.3} & {\bf 1.0} & {\bf 38.1} & {\bf 3.8} & {\bf 10.1} & {\bf 0.35} & {\bf 11} \\
WD0032-175 & 00:35:17.47 & -17:18:51.1 & 7.1 & 1.9 & 1.2 & 1.5 & 0.50 &  \\
WD0032-177 & 00:35:25.20 & -17:30:40.4 & 7.1 & 4.8 & 2.2 & 2.2 & 0.70 &  \\
WD0033+016 & 00:35:35.93 & +01:53:06.5 & 8.1 & 4.3 & 4.0 & 1.1 & 0.95 &  \\
{\bf WD0037-006} & {\bf 00:40:22.94} & {\bf -00:21:31.1} & {\bf 45.0} & {\bf 128.5} & {\bf 1.1} & {\bf 118.2} & {\bf 0.55} & {\bf 1} \\
HE0043-0318 & 00:46:18.38 & -03:02:00.8 & 45.0 & 2.1 & 1.0 & 2.1 & 0.45 &  \\
WD0047-524 & 00:50:03.74 & -52:08:17.1 & 48.9 & 0.3 & 0.8 & 0.4 & 0.65 &  \\
WD0048-544 & 00:51:08.87 & -54:11:21.2 & 48.9 & 1.1 & 2.0 & 0.6 & 0.60 &  \\
WD0048+202 & 00:51:11.00 & +20:31:22.3 & 73.9 & 2.9 & 2.4 & 1.2 & 0.70 &  \\
HE0049-0940 & 00:52:15.30 & -09:24:20.3 & 0.9 & 0.3 & 1.0 & 0.3 & 0.45 &  \\
WD0050-332 & 00:53:17.43 & -32:59:56.8 & 319.1 & 6.5 & 3.4 & 1.9 & 0.60 &  \\
WD0052-147 & 00:54:55.86 & -14:26:09.1 & 0.9 & 0.7 & 2.0 & 0.3 & 0.75 &  \\
WD0053-117 & 00:55:50.33 & -11:27:31.3 & 0.9 & 0.2 & 1.0 & 0.2 & 0.20 &  \\
HE0103-3253 & 01:05:30.77 & -32:37:54.3 & 10.1 & 3.3 & 1.3 & 2.4 & 0.60 &  \\
WD0103-278 & 01:05:53.52 & -27:36:56.8 & 3.9 & 5.5 & 0.8 & 6.8 & 0.50 &  \\
WD0106-358 & 01:08:20.75 & -35:34:43.0 & 4.0 & 1.6 & 2.3 & 0.7 & 0.60 &  \\
HE0106-3253 & 01:08:36.07 & -32:37:43.5 & 261.2 & 0.5 & 1.1 & 0.4 & 0.75 &  \\
WD0107-192 & 01:09:33.13 & -19:01:19.2 & 4.0 & 1.1 & 2.7 & 0.4 & 0.55 &  \\
WD0108+143 & 01:10:55.14 & +14:39:21.3 & 4.0 & 0.7 & 3.0 & 0.2 & 0.75 &  \\
WD0110-139 & 01:13:09.85 & -13:39:35.8 & 3.0 & 3.3 & 2.5 & 1.3 & 0.65 &  \\
{\bf WD0114-605} & {\bf 01:16:19.55} & {\bf -60:16:07.6} & {\bf 343.0} & {\bf 10.9} & {\bf 2.4} & {\bf 4.6} & {\bf 0.50} & {\bf } \\
WD0124-257 & 01:26:55.90 & -25:30:53.7 & 3.0 & 4.5 & 3.3 & 1.4 & 0.55 &  \\
WD0126+101 & 01:29:24.38 & +10:22:59.7 & 4.0 & 1.8 & 1.0 & 1.9 & 0.30 &  \\
WD0127-050 & 01:30:23.06 & -04:47:57.8 & 8.1 & 0.8 & 1.1 & 0.8 & 0.60 &  \\
WD0129-205 & 01:31:39.21 & -20:19:59.1 & 4.1 & 1.0 & 2.2 & 0.5 & 0.60 &  \\
HE0130-2721 & 01:33:09.08 & -27:05:45.0 & 12.1 & 2.2 & 2.2 & 1.0 & 0.55 &  \\
{\bf HE0131+0149} & {\bf 01:34:28.46} & {\bf +02:04:21.4} & {\bf 392.9} & {\bf 21.6} & {\bf 1.3} & {\bf 16.2} & {\bf 0.50} & {\bf } \\
WD0133-116 & 01:36:13.39 & -11:20:31.3 & 4.1 & 0.5 & 1.4 & 0.4 & 0.50 &  \\
{\bf WD0135-052} & {\bf 01:37:59.40} & {\bf -04:59:44.9} & {\bf 4.0} & {\bf 132.4} & {\bf 0.8} & {\bf 159.3} & {\bf 0.20} & {\bf 1, 5} \\
MCT0136-2010 & 01:38:31.67 & -19:54:50.6 & 6.1 & 0.9 & 1.5 & 0.6 & 0.70 &  \\
WD0137-291 & 01:40:16.79 & -28:52:53.7 & 2.9 & 6.1 & 1.8 & 3.4 & 0.50 &  \\
WD0140-392 & 01:42:50.99 & -38:59:06.9 & 2.0 & 1.1 & 1.2 & 0.9 & 0.55 &  \\
WD0145-221 & 01:47:21.76 & -21:56:51.4 & 2.1 & 6.5 & 2.0 & 3.3 & 0.60 &  \\
HS0145+1737 & 01:48:21.51 & +17:52:13.5 & 10.9 & 4.1 & 1.3 & 3.3 & 0.60 &  \\
HE0145-0610 & 01:48:22.27 & -05:55:36.5 & 0.0 & 9.9 & 3.8 & 2.6 & 0.45 &  \\
WD0151+017 & 01:54:13.88 & +02:01:23.5 & 3.0 & 0.5 & 1.4 & 0.4 & 0.50 &  \\
HE0152-5009 & 01:54:35.98 & -49:55:01.9 & 12.0 & 4.3 & 1.1 & 3.8 & 0.50 &  \\
WD0155+069 & 01:57:41.33 & +07:12:03.8 & 121.8 & 4.1 & 3.0 & 1.4 & 0.50 &  \\
HS0200+2449 & 02:03:45.80 & +25:04:09.1 & 120.7 & 0.6 & 2.0 & 0.3 & 0.60 &  \\
WD0204-233 & 02:06:45.10 & -23:16:14.0 & 2.0 & 0.1 & 1.0 & 0.1 & 0.45 &  \\
HE0204-4213 & 02:06:49.89 & -41:59:25.8 & 4.8 & 0.9 & 3.0 & 0.3 & 0.55 &  \\
WD0205-304 & 02:07:40.86 & -30:10:59.6 & 373.1 & 1.5 & 1.2 & 1.3 & 0.65 &  \\
HE0205-2945 & 02:08:08.00 & -29:31:38.8 & 357.0 & 0.5 & 2.7 & 0.2 & 0.35 &  \\
HE0210-2012 & 02:13:01.93 & -19:58:35.2 & 353.0 & 0.7 & 1.1 & 0.7 & 0.65 &  \\
HE0211-2824 & 02:13:56.66 & -28:10:17.8 & 61.9 & 1.1 & 1.0 & 1.1 & 0.55 &  \\
WD0212-231 & 02:14:21.26 & -22:54:49.1 & 79.0 & 10.0 & 3.8 & 2.6 & 0.65 &  \\
HE0219-4049 & 02:21:19.69 & -40:35:29.7 & 6.0 & 8.2 & 2.2 & 3.8 & 0.55 &  \\

\hline
\end{tabular}
\end{center}
\end{table*}

\begin{table*}
\contcaption{}
\begin{center}
\begin{tabular}{l r r r r r r r l}
\hline
\multicolumn{1}{l}{Name} &
\multicolumn{1}{l}{RA} &
\multicolumn{1}{l}{Dec} &
\multicolumn{1}{c}{$\Delta t$} &
\multicolumn{1}{c}{\drvm} & 
\multicolumn{1}{c}{$\sigma$} &
\multicolumn{1}{c}{$N_\sigma$} &
\multicolumn{1}{c}{$M_1$} &
\multicolumn{1}{l}{Comments} \\
& & &
\multicolumn{1}{c}{[d]} &
\multicolumn{1}{c}{[km\,s$^{-1}$]} &
\multicolumn{1}{c}{[km\,s$^{-1}$]} &
&
\multicolumn{1}{c}{[\msun]} & \\
\hline
{\bf HE0221-2642} & {\bf 02:23:29.40} & {\bf -26:29:19.7} & {\bf 226.3} & {\bf 14.0} & {\bf 5.9} & {\bf 2.4} & {\bf 0.55} & {\bf } \\
{\bf HE0221-0535} & {\bf 02:23:59.88} & {\bf -05:21:45.9} & {\bf 0.9} & {\bf 14.3} & {\bf 2.0} & {\bf 7.3} & {\bf 0.60} & {\bf } \\
{\bf HE0225-1912} & {\bf 02:27:41.43} & {\bf -18:59:24.5} & {\bf 7.1} & {\bf 48.1} & {\bf 3.1} & {\bf 15.5} & {\bf 0.55} & {\bf 1} \\
HS0225+0010 & 02:27:55.50 & +00:23:39.1 & 101.7 & 3.4 & 1.4 & 2.4 & 0.50 &  \\
WD0226-329 & 02:28:27.70 & -32:42:35.9 & 5.0 & 4.7 & 1.1 & 4.1 & 0.55 &  \\
WD0227+050 & 02:30:16.66 & +05:15:50.7 & 119.7 & 2.5 & 0.7 & 3.7 & 0.65 &  \\
WD0231-054 & 02:34:07.73 & -05:11:39.6 & 2.9 & 2.7 & 1.8 & 1.5 & 0.90 &  \\
HE0246-5449 & 02:48:07.16 & -54:36:44.9 & 1.9 & 0.8 & 1.6 & 0.5 & 0.60 &  \\
WD0250-026 & 02:52:51.05 & -02:25:17.4 & 70.9 & 1.3 & 1.5 & 0.9 & 0.50 &  \\
WD0250-007 & 02:53:32.29 & -00:33:45.3 & 18.9 & 1.5 & 2.8 & 0.5 & 0.40 &  \\
WD0252-350 & 02:54:37.25 & -34:49:56.6 & 1.1 & 3.2 & 1.0 & 3.1 & 0.45 &  \\
WD0255-705 & 02:56:16.90 & -70:22:17.7 & 2.0 & 3.2 & 3.4 & 1.0 & 0.50 &  \\
HE0256-1802 & 02:58:59.54 & -17:50:20.3 & 78.9 & 0.8 & 2.7 & 0.3 & 0.55 &  \\
HE0257-2104 & 02:59:52.65 & -20:52:49.6 & 78.9 & 0.8 & 2.0 & 0.4 & 0.55 &  \\
HE0300-2313 & 03:02:36.69 & -23:01:52.0 & 1.1 & 7.4 & 2.2 & 3.3 & 0.85 &  \\
WD0302+027 & 03:04:37.40 & +02:56:56.6 & 1.9 & 4.3 & 4.3 & 1.0 & 0.50 &  \\
HE0303-2041 & 03:06:04.96 & -20:29:31.1 & 6.1 & 1.4 & 1.8 & 0.8 & 0.55 &  \\
HE0305-1145 & 03:08:10.25 & -11:33:45.7 & 1.0 & 0.8 & 2.6 & 0.3 & 0.60 &  \\
WD0307+149 & 03:09:53.95 & +15:05:22.1 & 1.0 & 7.9 & 2.7 & 2.9 & 0.55 &  \\
HS0307+0746 & 03:10:09.13 & +07:57:32.6 & 135.7 & 2.6 & 2.7 & 1.0 & 0.50 &  \\
WD0310-688 & 03:10:30.99 & -68:36:03.3 & 74.8 & 1.6 & 0.5 & 3.2 & 0.70 &  \\
HE0308-2305 & 03:11:07.24 & -22:54:05.6 & 2.0 & 1.4 & 1.9 & 0.8 & 0.90 &  \\
WD0308+188 & 03:11:49.22 & +19:00:55.5 & 117.8 & 2.7 & 1.0 & 2.8 & 0.50 &  \\
HS0309+1001 & 03:12:34.96 & +10:12:27.2 & 198.3 & 8.9 & 2.9 & 3.1 & 0.45 &  \\
HS0315+0858 & 03:17:43.18 & +09:09:55.2 & 0.9 & 2.9 & 1.4 & 2.1 & 0.55 &  \\
HE0315-0118 & 03:18:13.31 & -01:07:13.1 & 2.0 & 7.4 & 2.3 & 3.2 & 0.50 & 1 \\
HE0317-2120 & 03:19:27.22 & -21:09:13.2 & 1.0 & 0.0 & 1.6 & 0.0 & 0.50 &  \\
WD0318-021 & 03:20:58.77 & -01:59:59.5 & 2.0 & 0.5 & 1.7 & 0.3 & 0.50 &  \\
{\bf HE0320-1917} & {\bf 03:22:31.91} & {\bf -19:06:47.8} & {\bf 1.0} & {\bf 70.4} & {\bf 1.2} & {\bf 61.1} & {\bf 0.30} & {\bf 10} \\
HE0324-2234 & 03:26:26.88 & -22:24:15.0 & 193.4 & 1.6 & 1.5 & 1.0 & 0.60 &  \\
HE0324-0646 & 03:26:39.97 & -06:36:05.2 & 5.0 & 4.5 & 1.1 & 4.2 & 0.65 &  \\
{\bf HE0324-1942} & {\bf 03:27:05.02} & {\bf -19:32:23.8} & {\bf 0.9} & {\bf 21.4} & {\bf 3.7} & {\bf 5.8} & {\bf 0.80} & {\bf 1} \\
{\bf HE0325-4033} & {\bf 03:27:43.92} & {\bf -40:23:26.1} & {\bf 0.9} & {\bf 26.8} & {\bf 1.5} & {\bf 17.4} & {\bf 0.55} & {\bf } \\
{\bf WD0326-273} & {\bf 03:28:48.81} & {\bf -27:19:01.7} & {\bf 3.0} & {\bf 179.3} & {\bf 1.2} & {\bf 151.2} & {\bf 0.35} & {\bf 3} \\
HE0330-4736 & 03:32:03.98 & -47:25:57.7 & 6.1 & 3.3 & 1.5 & 2.2 & 0.60 &  \\
WD0330-009 & 03:32:36.90 & -00:49:36.6 & 116.9 & 1.6 & 4.8 & 0.3 & 0.55 &  \\
HE0333-2201 & 03:36:02.77 & -21:51:21.5 & 34.9 & 2.7 & 1.0 & 2.8 & 0.70 &  \\
HE0336-0741 & 03:38:26.79 & -07:31:54.6 & 2.0 & 1.0 & 1.9 & 0.5 & 0.65 &  \\
WD0336+040 & 03:38:56.21 & +04:09:43.0 & 155.8 & 0.8 & 2.1 & 0.4 & 0.40 &  \\
HS0337+0939 & 03:39:58.55 & +09:49:11.3 & 134.8 & 3.0 & 1.9 & 1.6 & 0.50 &  \\
HE0338-3025 & 03:40:18.33 & -30:15:36.0 & 1.0 & 1.7 & 1.8 & 0.9 & 0.55 &  \\
{\bf WD0341+021} & {\bf 03:44:10.77} & {\bf +02:15:29.9} & {\bf 201.2} & {\bf 117.1} & {\bf 2.6} & {\bf 44.7} & {\bf 0.30} & {\bf } \\
{\bf WD0344+073} & {\bf 03:46:51.42} & {\bf +07:28:01.9} & {\bf 215.3} & {\bf 91.3} & {\bf 1.5} & {\bf 60.2} & {\bf 0.35} & {\bf } \\
HS0344+0944 & 03:46:52.31 & +09:53:56.1 & 134.8 & 8.9 & 2.4 & 3.7 & 0.75 &  \\
{\bf HE0344-1207} & {\bf 03:47:06.71} & {\bf -11:58:08.5} & {\bf 1.0} & {\bf 11.1} & {\bf 3.1} & {\bf 3.5} & {\bf 0.55} & {\bf } \\
HS0345+1324 & 03:48:39.58 & +13:33:29.3 & 155.7 & 4.3 & 2.8 & 1.6 & 0.75 &  \\
HS0346+0755 & 03:49:15.29 & +08:04:53.6 & 21.0 & 2.0 & 2.3 & 0.9 & 0.45 &  \\
HE0348-4445 & 03:49:59.27 & -44:36:27.4 & 2.1 & 1.0 & 2.6 & 0.4 & 0.70 &  \\
HE0348-2404 & 03:50:38.82 & -23:55:45.2 & 1.0 & 2.1 & 1.2 & 1.7 & 0.55 &  \\
HE0349-2537 & 03:51:41.37 & -25:28:16.6 & 34.9 & 4.8 & 3.7 & 1.3 & 0.55 &  \\
WD0352+052 & 03:54:41.09 & +05:23:19.4 & 196.3 & 2.1 & 3.2 & 0.6 & 0.45 &  \\
WD0352+018 & 03:54:43.47 & +01:58:41.4 & 206.2 & 0.4 & 3.7 & 0.1 & 0.55 &  \\
WD0352+096 & 03:55:22.02 & +09:47:17.5 & 214.2 & 0.6 & 1.1 & 0.6 & 0.70 &  \\
HE0358-5127 & 03:59:38.30 & -51:18:41.5 & 3.0 & 3.8 & 1.6 & 2.5 & 0.60 &  \\
HS0400+1451 & 04:03:42.08 & +14:59:28.9 & 201.2 & 2.7 & 1.7 & 1.6 & 0.85 &  \\
HE0403-4129 & 04:05:30.11 & -41:21:10.2 & 3.0 & 3.1 & 3.1 & 1.0 & 0.60 &  \\
WD0407+179 & 04:10:10.33 & +18:02:24.0 & 21.0 & 2.2 & 1.1 & 2.0 & 0.50 &  \\
WD0408-041 & 04:11:02.17 & -03:58:22.2 & 352.0 & 2.3 & 1.6 & 1.5 & 0.55 &  \\
HE0409-5154 & 04:11:10.33 & -51:46:50.8 & 363.0 & 8.5 & 3.0 & 2.8 & 0.55 &  \\
{\bf HE0410-1137} & {\bf 04:12:28.99} & {\bf -11:30:08.3} & {\bf 4.0} & {\bf 40.9} & {\bf 1.4} & {\bf 28.7} & {\bf 0.50} & {\bf 1} \\

\hline
\end{tabular}
\end{center}
\end{table*}

\begin{table*}
\contcaption{}
\begin{center}
\begin{tabular}{l r r r r r r r l}
\hline
\multicolumn{1}{l}{Name} &
\multicolumn{1}{l}{RA} &
\multicolumn{1}{l}{Dec} &
\multicolumn{1}{c}{$\Delta t$} &
\multicolumn{1}{c}{\drvm} & 
\multicolumn{1}{c}{$\sigma$} &
\multicolumn{1}{c}{$N_\sigma$} &
\multicolumn{1}{c}{$M_1$} &
\multicolumn{1}{l}{Comments} \\
& & &
\multicolumn{1}{c}{[d]} &
\multicolumn{1}{c}{[km\,s$^{-1}$]} &
\multicolumn{1}{c}{[km\,s$^{-1}$]} &
&
\multicolumn{1}{c}{[\msun]} & \\
\hline
WD0410+117 & 04:12:43.60 & +11:51:48.5 & 325.1 & 0.6 & 1.0 & 0.6 & 0.60 &  \\
HS0412+0632 & 04:14:58.36 & +06:40:07.0 & 333.1 & 1.2 & 0.9 & 1.3 & 0.55 &  \\
HE0414-4039 & 04:16:02.87 & -40:32:11.7 & 6.0 & 6.6 & 2.8 & 2.3 & 0.65 &  \\
WD0416-550 & 04:17:11.51 & -54:57:47.9 & 16.0 & 6.6 & 1.8 & 3.6 & 0.30 &  \\
HE0416-3852 & 04:18:04.14 & -38:45:20.6 & 6.0 & 2.3 & 2.1 & 1.1 & 0.75 &  \\
HE0416-1034 & 04:18:47.84 & -10:27:09.6 & 28.9 & 3.6 & 1.8 & 2.0 & 0.60 &  \\
{\bf HE0417-3033} & {\bf 04:19:22.07} & {\bf -30:26:44.0} & {\bf 261.2} & {\bf 10.2} & {\bf 3.1} & {\bf 3.3} & {\bf 0.50} & {\bf } \\
HE0418-5326 & 04:19:24.83 & -53:19:17.4 & 18.0 & 1.5 & 4.3 & 0.3 & 0.55 &  \\
HE0418-1021 & 04:21:12.03 & -10:14:09.0 & 260.2 & 2.6 & 2.3 & 1.1 & 0.80 &  \\
WD0421+162 & 04:23:55.81 & +16:21:13.9 & 307.1 & 2.0 & 1.8 & 1.1 & 0.70 &  \\
HE0425-2015 & 04:27:39.77 & -20:09:15.2 & 2.1 & 5.6 & 3.2 & 1.7 & 0.75 &  \\
WD0425+168 & 04:28:39.48 & +16:58:10.4 & 11.0 & 0.0 & 1.3 & 0.0 & 0.65 &  \\
HE0426-1011 & 04:28:42.32 & -10:04:48.9 & 202.3 & 3.7 & 1.4 & 2.7 & 0.70 &  \\
HE0426-0455 & 04:29:26.32 & -04:48:46.7 & 3.9 & 0.5 & 1.1 & 0.4 & 0.60 &  \\
WD0431+126 & 04:33:45.08 & +12:42:40.4 & 1.0 & 1.8 & 1.6 & 1.2 & 0.60 &  \\
HE0436-1633 & 04:38:47.33 & -16:27:21.4 & 225.3 & 2.6 & 1.0 & 2.5 & 0.60 &  \\
WD0437+152 & 04:39:52.97 & +15:19:44.0 & 4.0 & 2.0 & 1.7 & 1.2 & 0.40 &  \\
WD0446-789 & 04:43:46.67 & -78:51:50.2 & 3.0 & 0.4 & 1.0 & 0.4 & 0.50 &  \\
HE0452-3429 & 04:54:05.85 & -34:25:05.9 & 6.9 & 5.2 & 2.3 & 2.2 & 0.50 &  \\
HE0452-3444 & 04:54:23.69 & -34:39:48.7 & 32.8 & 0.5 & 1.7 & 0.3 & 0.55 &  \\
HE0456-2347 & 04:58:51.47 & -23:42:55.7 & 16.1 & 3.2 & 2.7 & 1.2 & 0.50 &  \\
HE0507-1855 & 05:09:20.47 & -18:51:17.3 & 2.0 & 0.1 & 2.8 & 0.0 & 0.80 &  \\
HS0507+0434B & 05:10:13.59 & +04:38:54.0 & 391.9 & 1.9 & 2.8 & 0.7 & 0.55 &  \\
HS0507+0434A & 05:10:14.01 & +04:38:37.4 & 399.8 & 1.3 & 1.3 & 1.0 & 0.55 &  \\
HE0508-2343 & 05:10:39.43 & -23:40:10.1 & 2.0 & 2.8 & 1.9 & 1.5 & 0.45 &  \\
WD0509-007 & 05:12:06.51 & -00:42:07.2 & 456.9 & 2.4 & 3.2 & 0.8 & 0.45 &  \\
WD0511+079 & 05:14:03.61 & +08:00:14.5 & 352.1 & 2.3 & 2.3 & 1.0 & 0.20 &  \\
{\bf HE0516-1804} & {\bf 05:19:04.27} & {\bf -18:01:29.1} & {\bf 1.0} & {\bf 22.3} & {\bf 2.7} & {\bf 8.2} & {\bf 0.55} & {\bf 13} \\
HE0532-5605 & 05:33:06.70 & -56:03:53.3 & 1.0 & 1.0 & 3.2 & 0.3 & 0.70 &  \\
WD0549+158 & 05:52:27.63 & +15:53:13.1 & 218.3 & 6.0 & 1.5 & 4.1 & 0.60 &  \\
WD0556+172 & 05:59:44.95 & +17:12:03.9 & 30.9 & 7.5 & 2.0 & 3.7 & 0.80 &  \\
WD0558+165 & 06:01:17.67 & +16:31:37.2 & 30.9 & 3.2 & 1.3 & 2.4 & 0.70 &  \\
WD0612+177 & 06:15:18.67 & +17:43:40.1 & 1.9 & 1.5 & 1.1 & 1.4 & 0.50 &  \\
WD0659-063 & 07:01:55.00 & -06:27:48.7 & 10.0 & 7.7 & 1.5 & 5.2 & 0.20 &  \\
WD0710+216 & 07:13:21.61 & +21:34:06.8 & 1.9 & 0.4 & 1.6 & 0.3 & 0.45 &  \\
WD0732-427 & 07:33:37.84 & -42:53:58.8 & 24.9 & 2.9 & 1.3 & 2.2 & 0.65 &  \\
WD0810-728 & 08:09:31.99 & -72:59:17.2 & 0.0 & 4.2 & 3.3 & 1.3 & 0.65 &  \\
WD0839-327 & 08:41:32.62 & -32:56:34.8 & 25.0 & 0.5 & 0.8 & 0.6 & 0.40 &  \\
WD0839+231 & 08:42:53.06 & +23:00:25.8 & 4.9 & 4.1 & 1.7 & 2.5 & 0.45 &  \\
WD0852+192 & 08:55:30.73 & +19:04:37.8 & 12.9 & 0.1 & 1.9 & 0.1 & 0.55 &  \\
WD0858+160 & 09:01:33.46 & +15:51:43.3 & 25.9 & 1.1 & 1.3 & 0.8 & 0.50 &  \\
WD0908+171 & 09:11:24.05 & +16:54:11.5 & 306.2 & 1.1 & 2.0 & 0.5 & 0.65 &  \\
WD0911-076 & 09:14:22.39 & -07:51:25.6 & 3.0 & 1.5 & 1.5 & 1.0 & 0.70 &  \\
WD0922+162A & 09:25:13.55 & +16:01:44.7 & 4.9 & 2.3 & 3.4 & 0.7 & 0.75 &  \\
WD0922+183 & 09:25:18.37 & +18:05:34.3 & 307.2 & 9.2 & 6.1 & 1.5 & 0.70 &  \\
WD0928-713 & 09:29:08.65 & -71:34:02.8 & 343.1 & 4.2 & 1.4 & 3.1 & 0.50 &  \\
HS0926+0828 & 09:29:36.53 & +08:15:46.8 & 334.0 & 6.7 & 7.3 & 0.9 & 0.55 &  \\
HS0929+0839 & 09:32:29.85 & +08:26:37.5 & 26.8 & 3.2 & 1.9 & 1.7 & 0.60 &  \\
HS0937+0130 & 09:39:58.67 & +01:16:38.2 & 280.2 & 2.8 & 3.0 & 0.9 & 0.90 &  \\
WD0937-103 & 09:40:11.96 & -10:34:25.1 & 4.9 & 7.4 & 1.5 & 5.0 & 0.90 &  \\
WD0939-153 & 09:41:56.22 & -15:32:14.6 & 3.0 & 0.1 & 1.1 & 0.1 & 0.55 &  \\
HS0940+1129 & 09:43:14.38 & +11:16:11.4 & 26.8 & 2.9 & 1.7 & 1.7 & 0.65 &  \\
HS0943+1401 & 09:46:31.60 & +13:47:35.8 & 31.8 & 8.6 & 3.7 & 2.3 & 0.75 &  \\
HS0944+1913 & 09:47:31.67 & +18:59:12.7 & 31.8 & 1.9 & 0.8 & 2.5 & 0.55 &  \\
HS0949+0935 & 09:51:48.94 & +09:21:12.6 & 63.9 & 6.3 & 3.1 & 2.0 & 0.65 &  \\
HS0949+0823 & 09:51:56.17 & +08:09:33.7 & 0.0 & 8.9 & 2.3 & 3.9 & 0.50 &  \\
WD0950+077 & 09:52:59.15 & +07:31:08.3 & 347.1 & 0.5 & 1.3 & 0.4 & 0.55 &  \\
WD0951-155 & 09:53:40.36 & -15:48:56.6 & 3.0 & 5.1 & 1.8 & 2.9 & 0.55 &  \\
WD0954+134 & 09:57:18.99 & +13:12:57.0 & 1.0 & 4.8 & 2.3 & 2.1 & 0.45 &  \\
WD0955+247 & 09:57:48.37 & +24:32:55.5 & 12.0 & 2.6 & 1.1 & 2.4 & 0.50 &  \\

\hline
\end{tabular}
\end{center}
\end{table*}

\begin{table*}
\contcaption{}
\begin{center}
\begin{tabular}{l r r r r r r r l}
\hline
\multicolumn{1}{l}{Name} &
\multicolumn{1}{l}{RA} &
\multicolumn{1}{l}{Dec} &
\multicolumn{1}{c}{$\Delta t$} &
\multicolumn{1}{c}{\drvm} & 
\multicolumn{1}{c}{$\sigma$} &
\multicolumn{1}{c}{$N_\sigma$} &
\multicolumn{1}{c}{$M_1$} &
\multicolumn{1}{l}{Comments} \\
& & &
\multicolumn{1}{c}{[d]} &
\multicolumn{1}{c}{[km\,s$^{-1}$]} &
\multicolumn{1}{c}{[km\,s$^{-1}$]} &
&
\multicolumn{1}{c}{[\msun]} & \\
\hline
WD0956+045 & 09:58:37.24 & +04:21:31.0 & 22.0 & 0.3 & 4.6 & 0.1 & 0.70 &  \\
WD0956+020 & 09:58:50.49 & +01:47:23.5 & 3.1 & 1.8 & 1.1 & 1.7 & 0.60 &  \\
WD1003-023 & 10:05:51.54 & -02:34:19.5 & 22.0 & 0.0 & 1.6 & 0.0 & 0.75 &  \\
HS1003+0726 & 10:06:23.08 & +07:12:12.6 & 238.1 & 5.8 & 4.4 & 1.3 & 0.50 &  \\
WD1010+043 & 10:13:12.78 & +04:05:12.8 & 1.0 & 9.9 & 4.4 & 2.2 & 0.60 &  \\
HE1012-0049 & 10:15:11.75 & -01:04:17.1 & 25.9 & 1.9 & 2.0 & 1.0 & 0.65 &  \\
HS1013+0321 & 10:15:48.15 & +03:06:46.8 & 245.1 & 1.5 & 1.8 & 0.8 & 0.55 &  \\
{\bf WD1013-010} & {\bf 10:16:07.01} & {\bf -01:19:18.7} & {\bf 20.9} & {\bf 29.0} & {\bf 2.3} & {\bf 12.6} & {\bf 0.25} & {\bf 12} \\
WD1015-216 & 10:17:26.67 & -21:53:43.4 & 25.8 & 2.9 & 2.6 & 1.1 & 0.60 &  \\
WD1015+161 & 10:18:03.84 & +15:51:58.3 & 4.9 & 0.5 & 1.0 & 0.5 & 0.65 &  \\
WD1017-138 & 10:19:52.45 & -14:07:35.5 & 269.0 & 4.2 & 2.8 & 1.5 & 0.60 &  \\
WD1017+125 & 10:19:56.02 & +12:16:29.9 & 15.1 & 5.9 & 2.2 & 2.7 & 0.55 &  \\
WD1019+129 & 10:22:28.77 & +12:41:59.4 & 15.0 & 3.6 & 1.1 & 3.2 & 0.65 &  \\
WD1020-207 & 10:22:43.83 & -21:00:02.1 & 25.8 & 2.7 & 1.2 & 2.3 & 0.65 &  \\
WD1026+023 & 10:29:09.87 & +02:05:49.7 & 408.9 & 0.4 & 1.0 & 0.4 & 0.65 &  \\
WD1031-114 & 10:33:42.79 & -11:41:40.4 & 2.0 & 1.6 & 1.0 & 1.7 & 0.55 &  \\
HS1043+0258 & 10:46:23.34 & +02:42:35.6 & 20.9 & 1.4 & 2.1 & 0.7 & 0.55 &  \\
WD1049-158 & 10:52:20.69 & -16:08:05.9 & 229.2 & 0.9 & 1.3 & 0.7 & 0.75 &  \\
WD1053-550 & 10:55:13.77 & -55:19:05.8 & 25.0 & 2.4 & 0.7 & 3.3 & 0.65 &  \\
WD1053-290 & 10:55:40.04 & -29:19:53.4 & 23.9 & 3.4 & 1.6 & 2.2 & 0.50 &  \\
HS1053+0844 & 10:55:51.54 & +08:28:46.6 & 4.9 & 3.8 & 1.9 & 2.0 & 0.65 &  \\
WD1056-384 & 10:58:20.19 & -38:44:26.5 & 25.0 & 0.6 & 1.4 & 0.4 & 0.60 &  \\
WD1058-129 & 11:01:12.28 & -13:14:42.7 & 166.7 & 2.3 & 2.6 & 0.9 & 1.00 &  \\
{\bf HS1102+0934} & {\bf 11:04:36.76} & {\bf +09:18:22.7} & {\bf 320.9} & {\bf 77.1} & {\bf 1.7} & {\bf 45.9} & {\bf 0.45} & {\bf 9} \\
WD1102-183 & 11:04:47.08 & -18:37:15.2 & 429.9 & 2.8 & 1.2 & 2.3 & 0.40 &  \\
HS1102+0032 & 11:05:15.33 & +00:16:26.3 & 264.0 & 4.0 & 2.9 & 1.4 & 0.80 &  \\
WD1105-048 & 11:07:59.98 & -05:09:27.3 & 269.1 & 3.1 & 0.6 & 5.1 & 0.50 &  \\
HS1115+0321 & 11:17:46.18 & +03:04:51.3 & 263.1 & 1.7 & 1.3 & 1.3 & 0.60 &  \\
WD1116+026 & 11:19:12.55 & +02:20:30.9 & 20.9 & 1.0 & 1.5 & 0.6 & 0.55 &  \\
HE1117-0222 & 11:19:34.66 & -02:39:06.3 & 25.0 & 6.3 & 1.0 & 6.4 & 0.60 &  \\
WD1121+216 & 11:24:13.08 & +21:21:34.8 & 38.9 & 0.0 & 1.0 & 0.0 & 0.25 &  \\
WD1122-324 & 11:24:35.62 & -32:46:25.7 & 25.9 & 0.2 & 1.7 & 0.1 & 0.60 &  \\
HE1124+0144 & 11:26:49.74 & +01:27:56.4 & 1.0 & 1.2 & 1.6 & 0.7 & 0.55 &  \\
WD1124-293 & 11:27:09.32 & -29:40:11.8 & 23.9 & 2.5 & 4.8 & 0.5 & 0.50 &  \\
{\bf WD1124-018} & {\bf 11:27:21.33} & {\bf -02:08:37.7} & {\bf 1.9} & {\bf 101.9} & {\bf 3.0} & {\bf 34.0} & {\bf 0.50} & {\bf } \\
WD1125-025 & 11:28:14.50 & -02:50:27.3 & 0.9 & 2.3 & 3.7 & 0.6 & 0.75 &  \\
WD1126-222 & 11:29:11.64 & -22:33:44.4 & 25.9 & 3.3 & 3.0 & 1.1 & 0.55 &  \\
WD1129+071 & 11:32:03.58 & +06:55:07.9 & 6.0 & 2.0 & 1.1 & 1.7 & 0.65 &  \\
WD1129+155 & 11:32:27.46 & +15:17:29.1 & 4.9 & 0.9 & 1.2 & 0.8 & 0.75 &  \\
WD1130-125 & 11:33:19.50 & -12:49:01.2 & 1.9 & 7.8 & 3.9 & 2.0 & 0.80 &  \\
HS1136+0326 & 11:39:26.64 & +03:10:19.7 & 0.9 & 5.6 & 2.2 & 2.6 & 0.55 &  \\
WD1144-246 & 11:47:20.13 & -24:54:56.7 & 24.9 & 1.4 & 2.1 & 0.7 & 0.35 &  \\
HS1144+1517 & 11:47:25.13 & +15:00:38.7 & 2.0 & 0.6 & 2.4 & 0.3 & 0.50 &  \\
WD1145+187 & 11:48:03.18 & +18:30:46.6 & 4.9 & 0.7 & 1.4 & 0.5 & 0.55 &  \\
WD1147+255 & 11:50:20.18 & +25:18:32.6 & 6.1 & 1.9 & 1.7 & 1.1 & 0.45 &  \\
WD1149+057 & 11:51:54.29 & +05:28:38.3 & 21.8 & 4.0 & 2.8 & 1.4 & 0.50 &  \\
WD1150-153 & 11:53:15.37 & -15:36:36.8 & 25.9 & 5.5 & 2.8 & 2.0 & 0.60 &  \\
HE1152-1244 & 11:54:34.91 & -13:01:16.8 & 27.0 & 0.5 & 0.9 & 0.6 & 0.50 &  \\
HS1153+1416 & 11:55:59.76 & +14:00:13.3 & 27.0 & 7.8 & 3.1 & 2.5 & 0.55 &  \\
WD1159-098 & 12:02:07.71 & -10:04:40.8 & 3.0 & 0.8 & 1.3 & 0.6 & 0.75 &  \\
WD1201-001 & 12:03:47.53 & -00:23:11.8 & 3.0 & 0.0 & 1.6 & 0.0 & 0.80 &  \\
WD1202-232 & 12:05:26.80 & -23:33:13.6 & 23.9 & 0.1 & 0.7 & 0.1 & 0.45 &  \\
WD1204-322 & 12:06:47.63 & -32:34:33.8 & 23.9 & 6.6 & 1.6 & 4.1 & 0.65 &  \\
WD1204-136 & 12:06:56.43 & -13:53:53.6 & 23.9 & 4.7 & 1.5 & 3.1 & 0.60 &  \\
{\bf HS1204+0159} & {\bf 12:07:29.51} & {\bf +01:42:50.6} & {\bf 1.0} & {\bf 11.3} & {\bf 4.2} & {\bf 2.7} & {\bf 0.50} & {\bf } \\
WD1207-157 & 12:10:09.34 & -16:00:40.4 & 3.0 & 0.3 & 1.2 & 0.2 & 0.55 &  \\
{\bf WD1210+140} & {\bf 12:12:33.89} & {\bf +13:46:25.1} & {\bf 0.9} & {\bf 133.1} & {\bf 1.8} & {\bf 72.6} & {\bf 0.30} & {\bf 4} \\
WD1216+036 & 12:18:41.15 & +03:20:21.7 & 11.0 & 0.0 & 1.5 & 0.0 & 0.50 &  \\
WD1220-292 & 12:23:05.17 & -29:32:28.9 & 1.1 & 0.9 & 1.1 & 0.8 & 0.70 &  \\
HE1225+0038 & 12:28:07.72 & +00:22:19.6 & 25.0 & 4.4 & 1.6 & 2.7 & 0.50 &  \\

\hline
\end{tabular}
\end{center}
\end{table*}

\begin{table*}
\contcaption{}
\begin{center}
\begin{tabular}{l r r r r r r r l}
\hline
\multicolumn{1}{l}{Name} &
\multicolumn{1}{l}{RA} &
\multicolumn{1}{l}{Dec} &
\multicolumn{1}{c}{$\Delta t$} &
\multicolumn{1}{c}{\drvm} & 
\multicolumn{1}{c}{$\sigma$} &
\multicolumn{1}{c}{$N_\sigma$} &
\multicolumn{1}{c}{$M_1$} &
\multicolumn{1}{l}{Comments} \\
& & &
\multicolumn{1}{c}{[d]} &
\multicolumn{1}{c}{[km\,s$^{-1}$]} &
\multicolumn{1}{c}{[km\,s$^{-1}$]} &
&
\multicolumn{1}{c}{[\msun]} & \\
\hline
WD1229-012 & 12:31:34.46 & -01:32:08.5 & 2.0 & 0.3 & 0.9 & 0.4 & 0.55 &  \\
WD1230-308 & 12:33:00.67 & -31:08:36.4 & 2.0 & 6.2 & 2.8 & 2.2 & 0.80 &  \\
WD1231-141 & 12:33:36.89 & -14:25:08.6 & 2.0 & 5.0 & 1.7 & 3.0 & 0.70 &  \\
{\bf WD1233-164} & {\bf 12:36:14.02} & {\bf -16:41:53.5} & {\bf 270.3} & {\bf 14.1} & {\bf 3.9} & {\bf 3.7} & {\bf 0.75} & {\bf } \\
WD1236-495 & 12:38:50.02 & -49:48:01.1 & 24.0 & 3.3 & 1.8 & 1.8 & 0.90 &  \\
WD1237-028 & 12:40:09.66 & -03:10:14.8 & 37.0 & 0.1 & 1.6 & 0.0 & 0.70 &  \\
WD1241-010 & 12:44:28.66 & -01:18:59.6 & 1.9 & 5.8 & 1.0 & 6.0 & 0.35 & 14 \\
HS1243+0132 & 12:45:38.74 & +01:16:16.1 & 263.1 & 2.6 & 5.4 & 0.5 & 0.60 &  \\
WD1244-125 & 12:47:26.88 & -12:48:42.0 & 42.8 & 1.7 & 1.2 & 1.4 & 0.60 &  \\
HE1247-1130 & 12:49:54.26 & -11:47:00.2 & 6.2 & 2.7 & 4.9 & 0.6 & 0.55 &  \\
HS1249+0426 & 12:52:15.19 & +04:10:43.0 & 33.0 & 7.6 & 2.6 & 2.9 & 0.50 &  \\
WD1249+160 & 12:52:17.15 & +15:44:43.4 & 1.8 & 1.9 & 1.3 & 1.4 & 0.30 &  \\
WD1249+182 & 12:52:23.34 & +17:56:53.9 & 1.0 & 2.3 & 2.6 & 0.9 & 0.55 &  \\
HE1252-0202 & 12:54:58.10 & -02:18:36.7 & 2.9 & 6.4 & 2.2 & 2.9 & 0.65 &  \\
WD1254+223 & 12:57:02.33 & +22:01:52.7 & 1.0 & 5.6 & 3.2 & 1.7 & 0.45 &  \\
WD1257+047 & 12:59:50.35 & +04:31:26.6 & 2.0 & 0.5 & 1.8 & 0.3 & 0.60 &  \\
WD1257+032 & 12:59:56.69 & +02:55:56.2 & 3.0 & 0.4 & 1.4 & 0.3 & 0.65 &  \\
HE1258+0123 & 13:01:10.50 & +01:07:39.9 & 24.1 & 0.6 & 2.6 & 0.2 & 0.45 &  \\
HE1307-0059 & 13:09:41.67 & -01:15:05.9 & 3.0 & 0.2 & 2.0 & 0.1 & 0.75 &  \\
HS1308+1646 & 13:11:06.06 & +16:31:03.4 & 26.0 & 0.2 & 3.7 & 0.0 & 0.60 &  \\
WD1308-301 & 13:11:17.52 & -30:25:57.6 & 15.9 & 0.1 & 0.8 & 0.1 & 0.55 &  \\
WD1310-305 & 13:13:41.59 & -30:51:33.7 & 1.0 & 1.9 & 1.6 & 1.2 & 0.65 &  \\
WD1314-153 & 13:16:43.59 & -15:35:58.7 & 1.0 & 2.4 & 1.3 & 1.9 & 0.55 &  \\
WD1314-067 & 13:17:18.46 & -06:59:28.1 & 1.2 & 3.1 & 1.7 & 1.9 & 0.55 &  \\
HE1315-1105 & 13:17:47.29 & -11:21:06.2 & 26.1 & 0.9 & 1.1 & 0.9 & 0.50 &  \\
WD1323-514 & 13:26:09.62 & -51:41:37.9 & 24.0 & 2.3 & 1.1 & 2.2 & 0.60 &  \\
HE1325-0854 & 13:28:23.90 & -09:09:53.0 & 2.0 & 2.2 & 0.8 & 2.6 & 0.55 &  \\
HE1326-0041 & 13:29:24.69 & -00:56:43.9 & 2.9 & 3.7 & 1.7 & 2.2 & 0.55 &  \\
WD1326-236 & 13:29:24.92 & -23:52:18.1 & 3.0 & 2.7 & 1.7 & 1.6 & 0.60 &  \\
WD1327-083 & 13:30:13.58 & -08:34:30.2 & 302.1 & 0.3 & 0.5 & 0.6 & 0.50 &  \\
WD1330+036 & 13:33:17.80 & +03:21:00.2 & 2.9 & 3.7 & 1.2 & 3.2 & 0.55 &  \\
WD1332-229 & 13:35:10.47 & -23:10:38.3 & 2.0 & 3.0 & 2.3 & 1.3 & 0.70 &  \\
{\bf HS1334+0701} & {\bf 13:36:33.67} & {\bf +06:46:26.8} & {\bf 309.2} & {\bf 13.1} & {\bf 2.0} & {\bf 6.5} & {\bf 0.40} & {\bf } \\
WD1334-160 & 13:36:59.29 & -16:19:44.1 & 3.0 & 0.7 & 1.3 & 0.5 & 0.80 &  \\
WD1334-678 & 13:38:08.11 & -68:04:37.4 & 289.0 & 5.0 & 2.5 & 2.0 & 0.40 &  \\
HE1335-0332 & 13:38:22.72 & -03:47:19.5 & 3.0 & 8.0 & 3.6 & 2.2 & 0.90 &  \\
HS1338+0807 & 13:41:27.63 & +07:52:29.5 & 59.9 & 0.3 & 2.7 & 0.1 & 0.50 &  \\
WD1342-237 & 13:45:46.58 & -23:57:11.0 & 3.0 & 8.3 & 2.2 & 3.9 & 0.55 &  \\
WD1344+106 & 13:47:24.45 & +10:21:36.6 & 311.0 & 0.8 & 1.1 & 0.8 & 0.20 &  \\
WD1348-273 & 13:51:22.84 & -27:33:59.1 & 1.0 & 4.4 & 3.2 & 1.4 & 0.45 &  \\
{\bf WD1349+144} & {\bf 13:51:54.06} & {\bf +14:09:44.2} & {\bf 1.1} & {\bf 84.3} & {\bf 2.7} & {\bf 30.7} & {\bf 0.55} & {\bf 1, 8} \\
WD1356-233 & 13:59:07.97 & -23:33:28.7 & 3.0 & 3.9 & 1.1 & 3.6 & 0.55 &  \\
WD1401-147 & 14:03:57.16 & -15:01:10.4 & 3.0 & 1.5 & 2.4 & 0.6 & 0.60 &  \\
WD1411+135 & 14:13:58.22 & +13:19:19.3 & 59.9 & 2.3 & 2.3 & 1.0 & 0.70 &  \\
WD1412-109 & 14:15:07.75 & -11:09:24.2 & 3.0 & 0.6 & 2.6 & 0.2 & 0.60 &  \\
HE1413+0021 & 14:16:00.21 & +00:07:59.3 & 2.9 & 0.6 & 1.7 & 0.3 & 0.65 &  \\
{\bf HE1414-0848} & {\bf 14:16:52.07} & {\bf -09:02:03.8} & {\bf 395.9} & {\bf 238.4} & {\bf 4.2} & {\bf 57.3} & {\bf 0.40} & {\bf 1, 2} \\
WD1418-088 & 14:20:54.82 & -09:05:08.7 & 3.0 & 2.5 & 1.5 & 1.7 & 0.45 &  \\
WD1420-244 & 14:23:26.25 & -24:43:29.4 & 2.9 & 4.4 & 2.2 & 2.0 & 0.80 &  \\
WD1422+095 & 14:24:39.24 & +09:17:12.7 & 3.0 & 0.5 & 1.6 & 0.3 & 0.55 &  \\
WD1426-276 & 14:29:27.38 & -27:51:01.3 & 0.9 & 1.9 & 1.3 & 1.5 & 0.60 &  \\
HS1430+1339 & 14:33:05.47 & +13:26:32.4 & 3.0 & 5.9 & 3.1 & 1.9 & 0.60 &  \\
WD1431+153 & 14:34:06.80 & +15:08:17.9 & 6.0 & 1.4 & 1.3 & 1.1 & 0.55 &  \\
HS1432+1441 & 14:35:20.85 & +14:28:41.3 & 5.0 & 2.5 & 1.7 & 1.5 & 0.50 &  \\
HE1441-0047 & 14:44:33.85 & -00:59:59.5 & 2.9 & 3.9 & 3.9 & 1.0 & 0.65 &  \\
HS1447+0454 & 14:50:09.91 & +04:41:45.7 & 59.0 & 3.1 & 1.1 & 2.8 & 0.55 &  \\
WD1448+077 & 14:50:49.46 & +07:33:32.9 & 323.0 & 3.6 & 1.1 & 3.1 & 0.55 &  \\
WD1449+168 & 14:52:11.37 & +16:38:03.5 & 6.0 & 3.0 & 1.7 & 1.7 & 0.50 &  \\
WD1451+006 & 14:53:50.48 & +00:25:29.3 & 2.9 & 2.7 & 1.8 & 1.5 & 0.55 &  \\
WD1457-086 & 14:59:52.99 & -08:49:29.5 & 2.9 & 1.9 & 2.0 & 0.9 & 0.55 &  \\

\hline
\end{tabular}
\end{center}
\end{table*}

\begin{table*}
\contcaption{}
\begin{center}
\begin{tabular}{l r r r r r r r l}
\hline
\multicolumn{1}{l}{Name} &
\multicolumn{1}{l}{RA} &
\multicolumn{1}{l}{Dec} &
\multicolumn{1}{c}{$\Delta t$} &
\multicolumn{1}{c}{\drvm} & 
\multicolumn{1}{c}{$\sigma$} &
\multicolumn{1}{c}{$N_\sigma$} &
\multicolumn{1}{c}{$M_1$} &
\multicolumn{1}{l}{Comments} \\
& & &
\multicolumn{1}{c}{[d]} &
\multicolumn{1}{c}{[km\,s$^{-1}$]} &
\multicolumn{1}{c}{[km\,s$^{-1}$]} &
&
\multicolumn{1}{c}{[\msun]} & \\
\hline
WD1500-170 & 15:03:14.45 & -17:11:56.7 & 3.0 & 5.2 & 3.3 & 1.5 & 0.65 &  \\
WD1501+032 & 15:04:23.92 & +03:02:30.5 & 3.0 & 0.5 & 1.0 & 0.5 & 0.55 &  \\
WD1503-093 & 15:06:19.44 & -09:30:20.9 & 5.0 & 1.3 & 1.7 & 0.7 & 0.60 &  \\
WD1507-105 & 15:10:29.08 & -10:45:19.8 & 5.0 & 5.1 & 2.2 & 2.3 & 0.30 &  \\
WD1511+009 & 15:14:21.31 & +00:47:52.3 & 3.0 & 4.1 & 3.1 & 1.3 & 0.55 &  \\
WD1515-164 & 15:18:35.07 & -16:37:29.2 & 3.0 & 1.1 & 1.3 & 0.8 & 0.70 &  \\
HS1517+0814 & 15:20:06.00 & +08:03:27.4 & 35.8 & 1.7 & 1.2 & 1.4 & 0.45 &  \\
HE1518-0344 & 15:20:46.03 & -03:54:52.2 & 1.0 & 0.2 & 3.9 & 0.1 & 0.60 &  \\
HE1518-0020 & 15:21:30.87 & -00:30:54.7 & 4.0 & 7.1 & 0.9 & 7.7 & 0.50 &  \\
HE1522-0410 & 15:25:12.26 & -04:21:29.3 & 16.9 & 7.2 & 3.0 & 2.4 & 0.55 &  \\
HS1527+0614 & 15:29:41.47 & +06:04:01.9 & 35.8 & 3.1 & 0.9 & 3.5 & 0.60 &  \\
WD1527+090 & 15:29:50.41 & +08:55:46.6 & 29.0 & 3.3 & 1.3 & 2.6 & 0.55 &  \\
WD1524-749 & 15:30:36.64 & -75:05:24.2 & 16.0 & 2.2 & 1.8 & 1.2 & 0.50 &  \\
WD1531-022 & 15:34:06.08 & -02:27:07.3 & 25.1 & 1.2 & 1.2 & 1.0 & 0.80 &  \\
WD1537-152 & 15:40:23.77 & -15:23:43.2 & 4.0 & 6.0 & 1.4 & 4.3 & 0.70 &  \\
WD1539-035 & 15:42:14.15 & -03:41:31.4 & 5.0 & 2.0 & 1.5 & 1.3 & 0.60 &  \\
WD1547+057 & 15:49:34.93 & +05:35:15.9 & 1.0 & 2.6 & 2.3 & 1.1 & 0.85 &  \\
WD1548+149 & 15:51:15.52 & +14:46:58.3 & 53.9 & 2.9 & 2.1 & 1.4 & 0.55 &  \\
WD1555-089 & 15:58:04.83 & -09:08:06.9 & 3.0 & 2.1 & 1.1 & 2.0 & 0.55 &  \\
WD1609+135 & 16:11:25.67 & +13:22:17.1 & 4.0 & 0.4 & 1.0 & 0.3 & 0.75 &  \\
WD1609+044 & 16:11:49.11 & +04:19:38.0 & 28.0 & 3.7 & 2.0 & 1.8 & 0.55 &  \\
HS1609+1426 & 16:12:06.51 & +14:19:05.8 & 48.0 & 3.1 & 1.3 & 2.5 & 0.50 &  \\
WD1614+136 & 16:16:52.31 & +13:34:21.5 & 0.9 & 0.6 & 0.9 & 0.6 & 0.30 &  \\
WD1614-128 & 16:17:28.02 & -12:57:45.6 & 1.8 & 0.1 & 1.3 & 0.1 & 0.60 &  \\
WD1615-154 & 16:17:55.24 & -15:35:52.7 & 1.8 & 2.5 & 1.5 & 1.6 & 0.70 &  \\
HS1616+0247 & 16:19:18.91 & +02:40:14.1 & 98.9 & 6.5 & 2.0 & 3.2 & 0.60 &  \\
WD1625+093 & 16:27:53.57 & +09:12:14.8 & 24.9 & 5.4 & 2.1 & 2.5 & 0.30 &  \\
WD1636+057 & 16:38:54.53 & +05:40:40.1 & 25.9 & 0.6 & 2.1 & 0.3 & 0.70 &  \\
WD1640+113 & 16:42:54.87 & +11:16:40.6 & 8.0 & 1.3 & 2.6 & 0.5 & 0.75 &  \\
HS1641+1124 & 16:43:54.12 & +11:18:50.2 & 0.9 & 2.6 & 1.6 & 1.6 & 0.60 &  \\
HS1646+1059 & 16:48:40.74 & +10:53:52.8 & 20.0 & 2.9 & 2.1 & 1.4 & 0.55 &  \\
HS1648+1300 & 16:51:02.78 & +12:55:12.7 & 1.0 & 0.8 & 1.4 & 0.6 & 0.50 &  \\
WD1655+215 & 16:57:09.84 & +21:26:48.4 & 20.0 & 0.7 & 1.3 & 0.5 & 0.45 &  \\
HS1705+2228 & 17:07:08.03 & +22:24:30.0 & 17.0 & 1.9 & 1.1 & 1.7 & 0.55 &  \\
WD1733-544 & 17:37:00.76 & -54:25:56.9 & 1.0 & 1.2 & 2.9 & 0.4 & 0.20 &  \\
WD1736+052 & 17:38:41.72 & +05:16:06.3 & 7.9 & 9.6 & 1.5 & 6.6 & 0.50 &  \\
WD1755+194 & 17:57:38.92 & +19:24:18.5 & 92.9 & 5.5 & 3.9 & 1.4 & 0.60 &  \\
WD1802+213 & 18:04:23.53 & +21:21:02.5 & 17.0 & 1.4 & 1.6 & 0.9 & 0.55 &  \\
{\bf WD1824+040} & {\bf 18:27:13.13} & {\bf +04:03:45.9} & {\bf 83.7} & {\bf 96.7} & {\bf 1.1} & {\bf 88.2} & {\bf 0.35} & {\bf 7} \\
WD1834-781 & 18:42:25.50 & -78:05:06.4 & 7.0 & 1.7 & 1.0 & 1.7 & 0.65 &  \\
WD1845+019 & 18:47:37.00 & +01:57:30.0 & 32.0 & 2.8 & 1.2 & 2.2 & 0.55 &  \\
WD1857+119 & 18:59:49.27 & +11:58:39.8 & 9.2 & 5.8 & 4.0 & 1.5 & 0.45 &  \\
WD1911+135 & 19:13:38.77 & +13:36:26.3 & 1.2 & 1.3 & 2.8 & 0.5 & 0.55 &  \\
WD1914-598 & 19:18:44.85 & -59:46:33.5 & 32.0 & 2.6 & 1.1 & 2.4 & 0.70 &  \\
WD1918+110 & 19:20:35.29 & +11:10:43.3 & 1.0 & 2.4 & 1.5 & 1.6 & 0.65 &  \\
WD1932-136 & 19:35:42.05 & -13:30:07.8 & 1.0 & 0.2 & 1.5 & 0.2 & 0.60 &  \\
WD1943+163 & 19:45:31.73 & +16:27:38.8 & 17.0 & 0.7 & 1.0 & 0.8 & 0.65 &  \\
WD1952-206 & 19:55:46.99 & -20:31:02.9 & 7.0 & 0.4 & 0.9 & 0.5 & 0.60 &  \\
WD1953-715 & 19:58:38.64 & -71:23:43.6 & 1.0 & 3.6 & 1.7 & 2.2 & 0.70 &  \\
WD1959+059 & 20:02:12.92 & +06:07:35.4 & 25.9 & 5.1 & 3.0 & 1.7 & 0.60 &  \\
WD2007-219 & 20:10:17.48 & -21:46:46.0 & 23.1 & 1.3 & 1.0 & 1.3 & 0.50 &  \\
WD2014-575 & 20:18:54.88 & -57:21:33.8 & 5.0 & 0.1 & 1.3 & 0.0 & 0.65 &  \\
WD2018-233 & 20:21:28.71 & -23:08:30.4 & 83.7 & 2.9 & 1.4 & 2.1 & 0.55 &  \\
{\bf WD2020-425} & {\bf 20:23:59.57} & {\bf -42:24:26.7} & {\bf 23.0} & {\bf 225.9} & {\bf 2.9} & {\bf 77.3} & {\bf 0.75} & {\bf 1} \\
WD2021-128 & 20:24:42.94 & -12:41:48.4 & 31.9 & 3.2 & 2.0 & 1.6 & 0.70 &  \\
WD2029+183 & 20:32:02.91 & +18:31:15.1 & 102.8 & 6.5 & 1.9 & 3.4 & 0.45 &  \\
WD2032+188 & 20:35:13.84 & +18:59:21.8 & 6.0 & 3.2 & 2.4 & 1.3 & 0.50 & 15 \\
WD2039-682 & 20:44:21.35 & -68:05:21.4 & 27.1 & 2.5 & 1.4 & 1.8 & 0.85 &  \\
{\bf HS2046+0044} & {\bf 20:48:38.26} & {\bf +00:56:00.8} & {\bf 16.9} & {\bf 23.8} & {\bf 3.8} & {\bf 6.2} & {\bf 0.70} & {\bf } \\
WD2046-220 & 20:49:46.18 & -21:54:43.1 & 83.7 & 1.0 & 1.8 & 0.6 & 0.50 &  \\

\hline
\end{tabular}
\end{center}
\end{table*}

\begin{table*}
\contcaption{}
\begin{center}
\begin{tabular}{l r r r r r r r l}
\hline
\multicolumn{1}{l}{Name} &
\multicolumn{1}{l}{RA} &
\multicolumn{1}{l}{Dec} &
\multicolumn{1}{c}{$\Delta t$} &
\multicolumn{1}{c}{\drvm} & 
\multicolumn{1}{c}{$\sigma$} &
\multicolumn{1}{c}{$N_\sigma$} &
\multicolumn{1}{c}{$M_1$} &
\multicolumn{1}{l}{Comments} \\
& & &
\multicolumn{1}{c}{[d]} &
\multicolumn{1}{c}{[km\,s$^{-1}$]} &
\multicolumn{1}{c}{[km\,s$^{-1}$]} &
&
\multicolumn{1}{c}{[\msun]} & \\
\hline
WD2051+095 & 20:53:43.18 & +09:41:14.5 & 25.8 & 1.6 & 1.8 & 0.8 & 0.50 &  \\
HS2056+0721 & 20:58:45.03 & +07:33:37.5 & 2.0 & 1.1 & 2.1 & 0.5 & 0.80 &  \\
WD2058+181 & 21:01:16.50 & +18:20:55.4 & 72.8 & 0.5 & 1.4 & 0.3 & 0.60 &  \\
HS2059+0208 & 21:01:47.77 & +02:20:27.6 & 17.9 & 1.5 & 2.5 & 0.6 & 0.65 &  \\
WD2059+190 & 21:02:02.68 & +19:12:57.5 & 1.0 & 0.3 & 1.8 & 0.2 & 0.20 &  \\
WD2115+010 & 21:17:33.58 & +01:15:47.1 & 1.0 & 4.7 & 1.9 & 2.5 & 0.50 &  \\
WD2122-467 & 21:25:30.19 & -46:30:36.8 & 58.9 & 3.0 & 1.9 & 1.5 & 0.75 &  \\
HS2132+0941 & 21:34:50.91 & +09:55:19.0 & 4.0 & 2.5 & 1.9 & 1.3 & 0.45 &  \\
HE2133-1332 & 21:36:16.18 & -13:18:33.0 & 30.0 & 2.8 & 1.0 & 2.9 & 0.35 &  \\
WD2134+218 & 21:36:36.15 & +22:04:32.8 & 61.8 & 3.9 & 1.3 & 3.1 & 0.60 &  \\
WD2136+229 & 21:38:46.21 & +23:09:20.9 & 69.8 & 4.1 & 2.2 & 1.9 & 0.50 &  \\
HE2135-4055 & 21:38:49.70 & -40:41:28.9 & 2.0 & 1.0 & 0.9 & 1.1 & 0.60 &  \\
WD2137-379 & 21:40:18.48 & -37:42:46.7 & 1.0 & 3.7 & 2.3 & 1.6 & 0.55 &  \\
HS2138+0910 & 21:41:03.02 & +09:23:45.4 & 4.0 & 4.1 & 1.1 & 3.8 & 0.40 &  \\
WD2139+115 & 21:41:28.37 & +11:46:22.1 & 3.1 & 0.3 & 1.2 & 0.2 & 0.60 &  \\
HE2140-1825 & 21:43:42.73 & -18:11:32.6 & 62.7 & 1.7 & 1.0 & 1.7 & 0.55 &  \\
HS2148+1631 & 21:51:14.54 & +16:45:23.1 & 1.0 & 1.1 & 1.7 & 0.6 & 0.55 &  \\
{\bf HE2148-3857} & {\bf 21:51:19.23} & {\bf -38:43:04.5} & {\bf 1.0} & {\bf 11.5} & {\bf 4.0} & {\bf 2.9} & {\bf 0.70} & {\bf } \\
WD2151-307 & 21:54:53.38 & -30:29:19.6 & 22.1 & 6.8 & 2.3 & 2.9 & 0.80 &  \\
HE2155-3150 & 21:58:46.08 & -31:36:06.5 & 59.7 & 4.3 & 1.6 & 2.6 & 0.65 &  \\
WD2157+161 & 21:59:34.35 & +16:25:39.0 & 1.0 & 0.8 & 2.5 & 0.3 & 0.55 &  \\
HE2159-1649 & 22:02:20.82 & -16:34:38.3 & 256.1 & 3.8 & 1.6 & 2.4 & 0.65 &  \\
{\bf WD2200-136} & {\bf 22:03:35.63} & {\bf -13:26:49.9} & {\bf 368.1} & {\bf 104.0} & {\bf 4.7} & {\bf 22.3} & {\bf 0.45} & {\bf 1} \\
WD2159-754 & 22:04:21.27 & -75:13:25.9 & 258.1 & 0.9 & 1.1 & 0.8 & 0.85 &  \\
HE2203-0101 & 22:06:02.44 & -00:46:33.5 & 399.0 & 6.0 & 1.6 & 3.7 & 0.65 &  \\
WD2204+071 & 22:07:16.20 & +07:18:36.0 & 3.1 & 3.9 & 2.4 & 1.6 & 0.65 &  \\
WD2207+142 & 22:09:47.19 & +14:29:46.6 & 5.9 & 1.0 & 2.0 & 0.5 & 0.35 &  \\
{\bf HE2209-1444} & {\bf 22:12:18.05} & {\bf -14:29:48.0} & {\bf 287.1} & {\bf 106.3} & {\bf 1.4} & {\bf 75.5} & {\bf 0.60} & {\bf 1, 6} \\
HS2210+2323 & 22:12:53.48 & +23:38:00.4 & 0.9 & 2.3 & 5.5 & 0.4 & 0.75 &  \\
{\bf HS2216+1551} & {\bf 22:18:57.15} & {\bf +16:06:56.9} & {\bf 1.1} & {\bf 12.4} & {\bf 1.9} & {\bf 6.4} & {\bf 0.65} & {\bf 1} \\
HE2218-2706 & 22:21:23.91 & -26:50:55.2 & 15.9 & 2.6 & 1.2 & 2.2 & 0.55 &  \\
HE2220-0633 & 22:22:44.44 & -06:17:54.9 & 397.0 & 0.3 & 1.5 & 0.2 & 0.60 &  \\
HS2220+2146B & 22:23:01.64 & +22:01:31.0 & 1.1 & 1.3 & 1.3 & 1.0 & 0.85 &  \\
HE2221-1630 & 22:24:17.51 & -16:15:47.0 & 290.3 & 0.1 & 1.5 & 0.0 & 0.55 &  \\
HS2225+2158 & 22:28:11.47 & +22:14:15.1 & 1.0 & 3.3 & 2.5 & 1.3 & 0.55 &  \\
WD2226+061 & 22:29:08.66 & +06:22:46.3 & 4.9 & 1.3 & 3.4 & 0.4 & 0.50 &  \\
WD2226-449 & 22:29:19.47 & -44:41:39.4 & 277.2 & 1.4 & 0.6 & 2.5 & 0.60 &  \\
HS2229+2335 & 22:31:45.45 & +23:51:23.9 & 1.0 & 3.9 & 1.5 & 2.6 & 0.70 &  \\
HE2230-1230 & 22:33:38.69 & -12:15:30.4 & 64.7 & 4.8 & 1.9 & 2.5 & 0.55 &  \\
HE2231-2647 & 22:34:02.59 & -26:32:21.1 & 233.3 & 2.8 & 2.4 & 1.2 & 0.50 &  \\
HS2233+0008 & 22:36:03.20 & +00:07:23.9 & 1.0 & 3.4 & 1.3 & 2.5 & 0.60 &  \\
WD2241-325 & 22:44:43.23 & -32:19:43.7 & 317.2 & 1.9 & 2.7 & 0.7 & 0.70 &  \\
HS2244+2103 & 22:46:45.28 & +21:19:47.7 & 0.9 & 3.1 & 3.4 & 0.9 & 0.60 &  \\
{\bf WD2248-504} & {\bf 22:51:02.02} & {\bf -50:11:31.8} & {\bf 7.0} & {\bf 10.1} & {\bf 2.8} & {\bf 3.6} & {\bf 0.60} & {\bf } \\
HE2251-6218 & 22:54:59.62 & -62:02:10.2 & 1.0 & 2.1 & 1.8 & 1.1 & 0.65 &  \\
{\bf WD2253-081} & {\bf 22:55:49.49} & {\bf -07:50:03.3} & {\bf 5.0} & {\bf 12.4} & {\bf 1.7} & {\bf 7.5} & {\bf 0.20} & {\bf } \\
{\bf WD2254+126} & {\bf 22:56:46.26} & {\bf +12:52:49.9} & {\bf 34.8} & {\bf 11.4} & {\bf 4.9} & {\bf 2.3} & {\bf 0.55} & {\bf } \\
HS2259+1419 & 23:01:55.18 & +14:36:00.5 & 1.0 & 1.8 & 1.1 & 1.6 & 0.55 &  \\
WD2303+242 & 23:06:17.70 & +24:32:07.5 & 1.0 & 0.6 & 1.7 & 0.4 & 0.50 &  \\
WD2306+130 & 23:08:30.58 & +13:19:22.7 & 370.0 & 3.9 & 1.5 & 2.6 & 0.60 &  \\
WD2306+124 & 23:08:35.07 & +12:45:39.0 & 39.9 & 5.9 & 2.1 & 2.8 & 0.70 &  \\
{\bf WD2308+050} & {\bf 23:11:18.05} & {\bf +05:19:27.9} & {\bf 1.0} & {\bf 11.4} & {\bf 4.3} & {\bf 2.7} & {\bf 0.45} & {\bf } \\
WD2312-356 & 23:15:34.95 & -35:24:51.8 & 18.2 & 3.8 & 1.1 & 3.6 & 0.55 &  \\
WD2314+064 & 23:16:50.36 & +06:41:27.6 & 0.9 & 1.0 & 2.8 & 0.4 & 0.60 &  \\
WD2318+126 & 23:20:31.30 & +12:58:14.5 & 0.9 & 1.3 & 1.6 & 0.8 & 0.60 &  \\
WD2322+206 & 23:24:35.22 & +20:56:33.9 & 0.9 & 2.1 & 1.1 & 1.9 & 0.50 &  \\
WD2322-181 & 23:25:18.40 & -17:51:57.8 & 3.0 & 2.9 & 1.4 & 2.0 & 0.65 &  \\
WD2324+060 & 23:26:44.55 & +06:17:41.4 & 0.9 & 1.3 & 1.2 & 1.1 & 0.60 &  \\
WD2326+049 & 23:28:47.74 & +05:14:53.5 & 42.0 & 5.3 & 1.1 & 5.0 & 0.55 &  \\
WD2328+107 & 23:30:41.79 & +11:02:05.0 & 2.9 & 0.7 & 1.3 & 0.5 & 0.55 &  \\

\hline
\end{tabular}
\end{center}
\end{table*}

\begin{table*}
\contcaption{}
\begin{center}
\begin{tabular}{l r r r r r r r l}
\hline
\multicolumn{1}{l}{Name} &
\multicolumn{1}{l}{RA} &
\multicolumn{1}{l}{Dec} &
\multicolumn{1}{c}{$\Delta t$} &
\multicolumn{1}{c}{\drvm} & 
\multicolumn{1}{c}{$\sigma$} &
\multicolumn{1}{c}{$N_\sigma$} &
\multicolumn{1}{c}{$M_1$} &
\multicolumn{1}{l}{Comments} \\
& & &
\multicolumn{1}{c}{[d]} &
\multicolumn{1}{c}{[km\,s$^{-1}$]} &
\multicolumn{1}{c}{[km\,s$^{-1}$]} &
&
\multicolumn{1}{c}{[\msun]} & \\
\hline
WD2329-332 & 23:32:10.90 & -33:01:08.1 & 3.1 & 1.2 & 3.5 & 0.3 & 0.60 &  \\
{\bf WD2330-212} & {\bf 23:32:59.48} & {\bf -20:57:12.1} & {\bf 2.9} & {\bf 55.5} & {\bf 2.6} & {\bf 21.8} & {\bf 0.40} & {\bf } \\
WD2333-165 & 23:35:36.59 & -16:17:42.5 & 42.0 & 1.7 & 0.6 & 2.6 & 0.50 &  \\
WD2333-049 & 23:35:53.96 & -04:42:14.8 & 40.9 & 2.6 & 2.7 & 1.0 & 0.50 &  \\
HE2334-1355 & 23:37:30.38 & -13:38:33.4 & 2.9 & 5.1 & 1.8 & 2.7 & 0.35 &  \\
{\bf WD2336-187} & {\bf 23:38:52.78} & {\bf -18:26:11.9} & {\bf 3.0} & {\bf 42.1} & {\bf 5.8} & {\bf 7.3} & {\bf 0.25} & {\bf 1} \\
WD2336+063 & 23:38:58.25 & +06:35:28.6 & 2.9 & 0.7 & 1.1 & 0.6 & 0.65 &  \\
MCT2343-1740 & 23:46:25.63 & -17:24:10.2 & 0.9 & 0.2 & 4.9 & 0.0 & 0.55 &  \\
MCT2345-3940 & 23:48:26.42 & -39:23:47.4 & 0.9 & 2.6 & 2.2 & 1.2 & 0.55 &  \\
WD2347+128 & 23:49:53.51 & +13:06:12.5 & 2.0 & 8.5 & 3.8 & 2.2 & 0.45 &  \\
WD2347-192 & 23:50:02.96 & -18:59:21.9 & 40.8 & 1.8 & 3.5 & 0.5 & 0.60 &  \\
HE2347-4608 & 23:50:32.90 & -45:51:34.8 & 10.2 & 2.3 & 1.3 & 1.9 & 0.45 &  \\
WD2348-244 & 23:51:22.10 & -24:08:17.0 & 39.9 & 0.2 & 1.7 & 0.1 & 0.50 &  \\
WD2349-283 & 23:52:23.18 & -28:03:15.9 & 39.9 & 2.1 & 1.1 & 2.0 & 0.50 &  \\
WD2350-248 & 23:53:03.79 & -24:32:03.1 & 41.7 & 2.5 & 2.9 & 0.9 & 0.85 &  \\
WD2350-083 & 23:53:27.63 & -08:04:39.5 & 63.9 & 4.0 & 1.4 & 2.9 & 0.55 &  \\
WD2351-368 & 23:54:18.82 & -36:33:55.1 & 4.0 & 1.6 & 1.6 & 1.0 & 0.55 &  \\
WD2354-151 & 23:57:33.44 & -14:54:09.1 & 41.8 & 4.7 & 2.4 & 1.9 & 0.35 &  \\
HE2356-4513 & 23:58:57.83 & -44:57:13.5 & 0.0 & 2.6 & 1.2 & 2.2 & 0.55 &  \\

\hline
\end{tabular}
\end{center}

\begin{flushleft}
Notes:
$\sigma$ is the root of the summed squares of the RV errors of the two individual RV measurements forming each difference.
$N_\sigma$ is defined as \drvm$/\sigma$.
$M_1$ is the derived mass for the photometric-primary WD.
DWD candidates with \drvm > 10\kms are highlighted.

(1)~Double-lined DWD
(2)~HE1414-0848: $P=0.5178$\,d, $M_1=0.55$\,\msun, $M_2=0.71$\,\msun\ \citep{Napiwotzki_2002}
(3)~WD0326-273: $P=1.8754$\,d, $M_1=0.51$\,\msun, $M_{2,\textrm{min}}=0.59$\,\msun\ \citep{Nelemans_2005}
(4)~WD1210+140: $P=0.64194$\,d, $M_1=0.23$\,\msun, $M_{2,\textrm{min}}=0.38$\,\msun\ \citep{Nelemans_2005}
(5)~WD0135-052: $P=1.553$\,d, $M_1=0.47$\,\msun, $M_2=0.52$\,\msun\ \citep{Saffer_1988,Bergeron_1989}
(6)~HE2209-1444: $P=0.2769$\,d, $M_1=0.58$\,\msun, $M_2=0.58$\,\msun\ \citep{Karl_2003b}
(7)~WD1824+040: $P=6.26600$\,d, $M_1=0.428$\,\msun, $M_{2,\textrm{min}}=0.515$\,\msun\ \citep{MoralesRueda_2005}
(8)~WD1349+144: $P=2.2094$\,d, $M_1=0.44$\,\msun, $M_2=0.44$\,\msun\ \citep{Karl_2003a}
(9)~HS1102+0934: $P=0.55319$\,d, $M_1=0.46$\,\msun, $M_{2,\textrm{min}}=0.55$\,\msun\ \citep{Brown_2013}
(10)~HE0320-1917: $P=0.86492$\,d, $M_1=0.29$\,\msun, $M_{2,\textrm{min}}=0.35$\,\msun\ \citep{Nelemans_2005}
(11)~WD0032-317: possible 1500\,K BD companion, based on weak NIR excess.
(12)~WD1013-010: $P=0.43653$\,d, $M_1=0.44$\,\msun, $M_{2,\textrm{min}}=0.38$\,\msun\ \citep{Nelemans_2005}
(13)~HE0516-1804: possible NIR excess indicating a $T_{\rm eff}\sim 3000$\,K companion with radius $\sim 0.2$\,\rsun.
(14)~WD1241-010: $P=3.34741$\,d, $M_1=0.31$\,\msun, $M_{2,\textrm{min}}=0.373$\,\msun\ \citep{Marsh_1995}
(15)~WD2032+188: $P=5.0846$\,d, $M_1=0.406$\,\msun, $M_{2,\textrm{min}}=0.469$\,\msun\ \citep{MoralesRueda_2005}

(A full version of this table appears in the electronic version of the article.)
\end{flushleft}
\end{table*}


%%%%%%%%%%%%%%%%%%%%%%%%%%%%%%%%%%%%%%%%%%%%%%%%%%

% Don't change these lines
\bsp	% typesetting comment
\label{lastpage}
\end{document}